\documentclass[
 reprint,
 amsmath,amssymb,
 aps,
]{revtex4-1}

\usepackage{graphicx}
\usepackage{dcolumn}
\usepackage{bm}
\renewcommand{\vec}[1]{\mathbf{#1}}
\usepackage{amsmath}
\usepackage{amsfonts}
\usepackage{amssymb}
\usepackage{dsfont}
\usepackage{tikz}
\usepackage{wrapfig}
\usepackage{braket}
\usepackage[caption=false]{subfig}
\usepackage{float}
\usepackage{placeins}
\usepackage{epstopdf}
\usepackage{hyperref}

\begin{document}

\preprint{APS/123-QED}

\title{Decay-dephasing-induced steady states in bosonic Rydberg-excited quantum gases in an optical lattice}

\author{Mathieu Barbier}
\email{barbier@th.physik.uni-frankfurt.de}
\author{Andreas Gei\ss{}ler}
\author{Walter Hofstetter}
\affiliation{Institut f\"ur Theoretische Physik, Goethe-Universit\"at, 60438 Frankfurt/Main, Germany
}

\date{\today}

\begin{abstract}
We investigate the possibility of realizing supersolid quantum phases in bosonic Rydberg-excited quantum lattice gases in the presence of non-unitary processes, by simulating the dynamical evolution starting from initial preparation in non-dissipative equilibrium states. Within Gutzwiller theory, we first analyze the many-body ground-state of a bosonic Rydberg-excited quantum gas in a two dimensional optical lattice for variable atomic hopping rates and Rabi detunings. Furthermore, we perform time evolution of different supersolid phases using the Lindblad-master equation. With the inclusion of two different non-unitary processes, namely spontaneous decay from a Rydberg state to the ground state and dephasing of the addressed Rydberg state, we study the effect of non-unitary processes on those quantum phases and observe long-lived states in the presence of decay and dephasing. We find that long-lived supersolid quantum phases are observable within a range of realistic decay and dephasing rates, while high rates cause any initial configuration to homogenize quickly, preventing possible supersolid formation.
\end{abstract}

\maketitle

\section{Introduction}
In recent years, numerous theoretical works have predicted exotic phases in Rydberg-excited bosonic quantum gases arising from the interplay between condensation and long-range spatial order. Alongside insulating density waves with crystalline order \cite{RydbergDensityWave,ScalesI} and superfluids \cite{RydbergSuperfluidI,RydbergSuperfluidII}, quantum phases referred to as supersolids have been predicted \cite{RydbergAndreas,RydbergSupersolidsI,RydbergSupersolidsII,RydbergSupersolidsIII,RydbergSupersolidsIV}. Those phases are characterized by a spatially modulated condensate combining the spontaneous long-range spatial order of solids and the superfluid flow of interacting condensates. While recent experiments strive towards the observation of such quantum phases, their realization still proves to be a challenge \cite{Observation, ProbableObservation}.\\
Several non-negligible, non-unitary processes turn out to be major obstacles in experiments, such as the spontaneous decay of a Rydberg-excited particle back to the ground state caused by the finite lifetime of Rydberg excitations \cite{RadiativeLifeI,RadiativeLifeII,RadiativeLifeIII,DissipationI,AllOptical} which effectively lowers the Rydberg fraction $n_e$ in the system, the dephasing of the Rydberg state, caused by blackbody-radiation induced decay to unadressed Rydberg levels \cite{DephasingI,DephasingII} and the finite linewidth of the laser \cite{Bistability,LamourPhD}, and the collective loss of particles \cite{AvalancheI,AtomLossII}.\\
A theoretical tool for the inclusion of non-unitary processes in the dynamics is the Markovian master equation in Lindblad form, used for the description of non-unitary density matrix evolution. In combination with the Gutzwiller approximation, assuming the decoupling of the many-body wave function into products of single-site wave functions, we will be able to simulate the time evolution, observe the effects of the decay and dephasing, and investigate whether or not stable supersolid quantum phases are attainable.

\section{Model and Theory}
For the description of a bosonic quantum gas with two atomic levels, i.e. a hyperfine ground state and a Rydberg-excited state, a multi-component extension of the single band Bose-Hubbard model is required. This Hamiltonian consists of three distinct parts: Two intraspecies terms and one interspecies term. Considering the ground state $|g\rangle$ and a Rydberg-excited state $|e\rangle$ of an atom, the Hamiltonian reads
\begin{equation}
\hat{H} = \hat{H}_g + \hat{H}_e + \hat{H}_{g,e}. \label{eq:Hamiltfull}
\end{equation}
Within second quantization, the Hamiltonian is written in Fock space in terms of creation and annihilation operators $(\hat{b}^{\nu}_i)^\dag$ and $\hat{b}^{\nu}_{i}$ for state $\nu \in \{g,e\}$ acting on site $i$. The Hamiltonian of the ground state atoms reads
\begin{equation}
\hat{H}_g = -\mu \sum_i \hat{n}^g_{i} - t_g \sum_{<ij>} (\hat{b}^g_{i})^\dag \hat{b}^g_j + \frac{1}{2}U_g\sum_i \hat{n}^g_{i}(\hat{n}^g_{i}-1)
\end{equation}
with the occupation number operator $\hat{n}^\nu_{i} = (\hat{b}^\nu_{i})^\dag \hat{b}^\nu_{i}$ and $<ij>$ denoting pairs of next-neighbor sites. This Hamiltonian is identical to the Bose-Hubbard model, in which we introduce the chemical potential $\mu$, the ground state hopping rate $t_g$ and the ground state on-site interaction $U_g$. It describes the physics of ground state atoms accurately \cite{ElectronCorrelations}. The Hamiltonian of the Rydberg excitations resembles its ground state counterpart, but contains additional terms.
\begin{equation}
\begin{split}
\hat{H}_e = &-\mu\sum_i \hat{n}^e_i - t_e \sum_{<ij>} (\hat{b}^e_i)^\dag \hat{b}^e_j + \frac{1}{2}U_e\sum_i \hat{n}^e_i(\hat{n}^e_i-1)\\
&-\Delta \sum_i \hat{n}^e_i + V\sum_i \sum_{j \neq i} \frac{\hat{n}^e_i\hat{n}^e_j}{|\vec{x_i}-\vec{x_j}|^6}
\end{split}
\end{equation}
The second to last term describes the effect of the detuning of the excitation laser and arises within the rotating wave approximation (RWA) \cite{RWAI}. Note that we consider identical trapping for ground state and Rydberg-excited atoms. This is a non-trivial case, since Rydberg excitations are in general antitrapped by the optical lattice, but recently there have been several proposals about the engineering of optical lattices trapping ground state and Rydberg-excited particles \cite{RydbergTrapI,RydbergTrapII}.\\
The last term corresponds to the van-der-Waals interaction of the Rydberg excitations. Highly excited atoms possess a high polarizability, inducing an interaction between pairs of atoms on different lattice sites. Finally, we have the interspecies Hamiltonian
\begin{equation}
\hat{H}_{g,e} = U_{ge} \sum_i \hat{n}^g_i \hat{n}^e_i + \frac{\Omega}{2} \sum_i ((\hat{b}^g_i)^\dag \hat{b}^e_i + (\hat{b}^e_i)^\dag \hat{b}^g_i)
\end{equation}
with an on-site interaction $U_{ge}$ and the Rabi driving with frequency $\Omega$ mixing the ground and excited states. This Rabi term arises alongside the detuning term within the RWA, and its frequency corresponds to an effective Rabi frequency within a two-photon excitation scheme \cite{RamanTransitions}. We do not include the long-range interaction between a Rydberg excitation and a ground state particle, since the interaction strength is negligible due to the low polarizability of the ground state particle \cite{Polarizability}. While most parameters of the Hamiltonian are rates, which can be experimentally well adjusted with high tunability and are thus chosen to fit the experimental setups, some have to be chosen in order to model observed physical phenomena. First, we impose a hardcore-like constraint $U_{g,e} \gg U_g$ motivated by the  energy scale of the interaction between a Rydberg excitation and another atom on short distances. Additionally, setting $U_e~\rightarrow \infty$ models an on-site Rydberg blockade: The suppression of excitation to the Rydberg state due to the frequency shift of the adressed Rydberg level of an already excited atom \cite{RydbergBlockadeI,RydbergBlockadeII}. Furthermore, we consider a negligible hopping rate of the Rydberg excitations, hence $t_e = 0$. This assumption is motivated by the overwhelming strength of the van-der-Waals interaction, which dominates all other energy scales. Due to the strong coupling of the Rydberg-excited state to the hyperfine ground state, which has finite hopping, the excitations may nevertheless become delocalized.\\
\par
With the full Hamiltonian at hand, several methods for its analysis exist. One of them is the Gutzwiller approximation (GA), in which we approximate the many-body state by a variational ansatz \cite{GutzwillerI,GutzwillerII,GutzwillerIII,GutzwillerIV}, which factorizes the wave function of the complete system into partial wave functions
\begin{equation}
|\Psi^{GA}\rangle = \Pi_i |\Psi\rangle_i \label{eq:product}
\end{equation}
and fulfills the Schr\"odinger equation $\hat{H}^{GA} |\Psi^{GA}\rangle = E^0 |\Psi^{GA}\rangle$. Note that the Hamiltonian $\hat{H}^{GA}$ is not equal to the original Hamiltonian $\hat{H}$, as we will discuss in detail below. Each $|\Psi\rangle_i$ describes the wave function on a site $i$ of the optical lattice. Motivated by the factorization of the wave function, we decouple the Hamiltonian $\hat{H}^{GA}$ into single site Hamiltonians $\hat{H}_i$, with the respective Schr\"odinger equations
\begin{equation}
\hat{H}_i |\Psi\rangle_i = E^0_i |\Psi\rangle_i
\end{equation}
with $\hat{H}^{GA} = \sum_i \hat{H}_i$ and $E^0 = \sum_i E^0_i$. This reduces the calculation of the ground-state via exact diagonalization. The original Hamiltonian $\hat{H}$ includes terms entangling different sites, namely the hopping mechanism and the van-der-Waals interaction.
We decouple these terms with the approximations
\begin{equation}
\begin{split}
(\hat{b}^\nu_i)^\dag \hat{b}^\nu_j &\approx (\hat{b}^\nu_i)^\dag \langle \hat{b}^\nu_j \rangle + \langle (\hat{b}^\nu_i)^\dag \rangle \hat{b}^\nu_j - \langle (\hat{b}^\nu_i)^\dag \rangle \langle \hat{b}^\nu_j \rangle,\\
\hat{n}^e_i \hat{n}^e_j &\approx \hat{n}^e_i \langle \hat{n}^e_j \rangle + \langle \hat{n}^e_i \rangle \hat{n}^e_j - \langle \hat{n}^e_i \rangle \langle \hat{n}^e_j \rangle,
\end{split}
\end{equation}
neglecting higher-order fluctuations $(\delta \hat{A})^n \approx 0$ with $\delta\hat{A} = \hat{A} - \langle \hat{A} \rangle$ and $\hat{A} \in \{\hat{b}, \hat{n}\}$ for $n \geq 2 $. Both are static mean-field approximations, the latter being the Hartree approximation. We allow a spatially inhomogeneous system due to the long-range ordering induced by the van-der-Waals interaction between Rydberg excitations.
\begin{figure}[t]
\begin{tabular}{c}
\subfloat[Checkerboard-structured primitive cell]{\includegraphics[width=0.33\textwidth, trim = 0 0 0 0, clip]{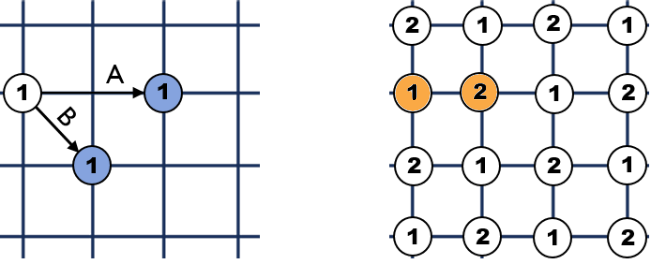}}\\
\subfloat[Triangle-structured primitive cell]{\includegraphics[width=0.33\textwidth, trim = 0 0 0 0, clip]{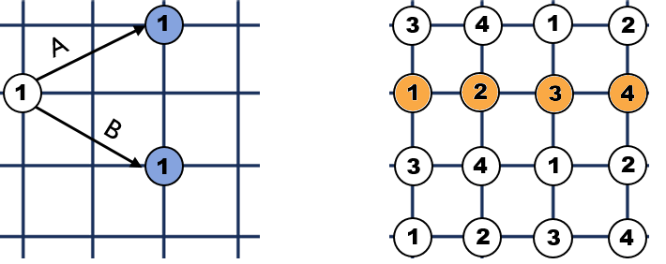}}\\
\subfloat[Arbitrarily structured primitive cell]{\includegraphics[width=0.33\textwidth, trim = 0 0 0 0, clip]{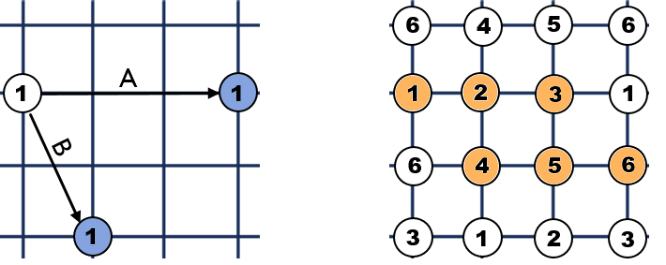}}\\
\end{tabular}
\caption{Various primitive cells, corresponding to different ordered structures: The spanning vectors $\vec{A}$ and $\vec{B}$ starting at an arbitrary site $i$ point to the next equivalent site, thus creating the boundaries of the primitive cell (colored in orange). The sites within are labeled by increasing numbers, going from left to right, up to down.} \label{fig:structures}
\end{figure}

Applying the above approximations to the original Hamiltonian \eqref{eq:Hamiltfull}, we obtain for each lattice site the single-site Hamiltonian
\begin{equation}
\hat{H}_i = \hat{H}^g_i + \hat{H}^e_i + \hat{H}^{g,e}_i \label{eq:HamiltApprox}
\end{equation}
with
\begin{equation}
\hspace*{-0.3cm} \hat{H}^g_i = -\mu \hat{n}^g_{i} - t_g ((\hat{b}^g_{i})^\dag \xi^g_{i} + \hat{b}^g_{i} (\xi^g_{i})^*) + \frac{1}{2}U_g \hat{n}^g_{i}(\hat{n}^g_{i}-1)
\end{equation}
\vspace*{-0.8cm}
\begin{align}
\hspace*{-0.75cm} \hat{H}^e_i = &-\mu \hat{n}^e_i - t_e ((\hat{b}^e_{i})^\dag \xi^e_{i} + \hat{b}^e_{i} (\xi^e_{i})^*) + \frac{1}{2}U_e \hat{n}^e_i(\hat{n}^e_i-1) \nonumber \\
\hspace*{-0.75cm} &-\Delta \hat{n}^e_i + V (\hat{n}^e_{i} \eta_i - \frac{1}{2} \langle \hat{n}^e_{i} \rangle \eta_i),
\end{align}
\vspace*{-0.8cm}
\begin{equation}
\hspace*{-2.5cm} \hat{H}^{g,e}_i = \text{ }U_{ge} \hat{n}^g_i \hat{n}^e_i + \frac{\Omega}{2} ((\hat{b}^g_i)^\dag \hat{b}^e_i + (\hat{b}^e_i)^\dag \hat{b}^g_i),
\end{equation}
where $\xi_i = \sum_{j.n.n.i} \langle \hat{b}_j \rangle$ (with $j.n.n.i$ denoting $j$ next neighbor to fixed $i$) and $\eta_i = \sum_{j \neq i} 2 \langle \hat{n}_j \rangle /|\vec{x}_i - \vec{x}_j|^6$ are arising from the mean-field approximation.\\
Solving the Schr\"odinger equation for the above single-site Hamiltonians involves a greatly reduced Hilbert space. However, non-local expectation values of the observables appear in each Hamiltonian via $\xi^\nu_i$ and $\eta_i$, effectively coupling different lattice sites.

\section{Gutzwiller ground-state phase diagram}

With the product ansatz \eqref{eq:product}, the many-body ground-state is given by the ground-state wave functions of all sites. Due to the coupling of the single-site Hamiltonians to the states of all the other sites, the single-site ground-states have to be obtained self-consistently. We assign initial states $|\Psi\rangle_i$ to each site and calculate the expectation values $\phi^g_i$,$\phi^e_i$,$n^g_i$ and $n^e_i$ (where $\phi^\nu_i = \langle \Psi |_i \hat{b}^\nu_i | \Psi \rangle_i$ is the local $\nu$-state condensate parameter of site $i$ and $n^\nu_i = \langle \Psi |_i \hat{n}^\nu_i | \Psi \rangle_i$ is the local $\nu$-state occupation number of site $i$). By performing updates on randomly selected sites, which consist of constructing the Hamiltonian of the selected site through the expectation values of all other sites and replacing the current state with the new eigenstate obtained via subsequent diagonalization, the single-site states converge to their respective ground-states and yield the self-consistent many-body ground-state of the whole system.
\par
\noindent
In the presence of periodic boundaries on a finite system, the spatial distribution of the Rydberg excitations is limited to certain crystalline structures compatible with the geometry of the periodic cell. Therefore, various primitive cells have to be tested in order to find the structure corresponding to the true ground-state. We perform the above mentioned iterative procedure on a number of primitive cells with varying structures and compare the resulting ground-state energies. The area of each primitive cell is given by the number of individual sites $N$. After a single iteration step, not only the randomly chosen site is updated, but also all equivalent sites labeled by the same number (see Fig. \ref{fig:structures}). Primitive cells are characterized by two spanning vectors $\vec{A}$ and $\vec{B}$, whose variation leads to different primitive structures, e.g. square or quasi-triangular or -hexagonal. Note that those primitive cells are created on a square 2D optical lattice, in which many regular non-square structures can only be approximated. For the true ground-state, we select the primitive cell yielding the lowest ground-state energy.

\begin{figure}[H]
\subfloat{    \begin{picture}(260,170)
    \put(0,0){\includegraphics[width=0.5\textwidth,trim = {0 0 0 1.5cm}, clip]{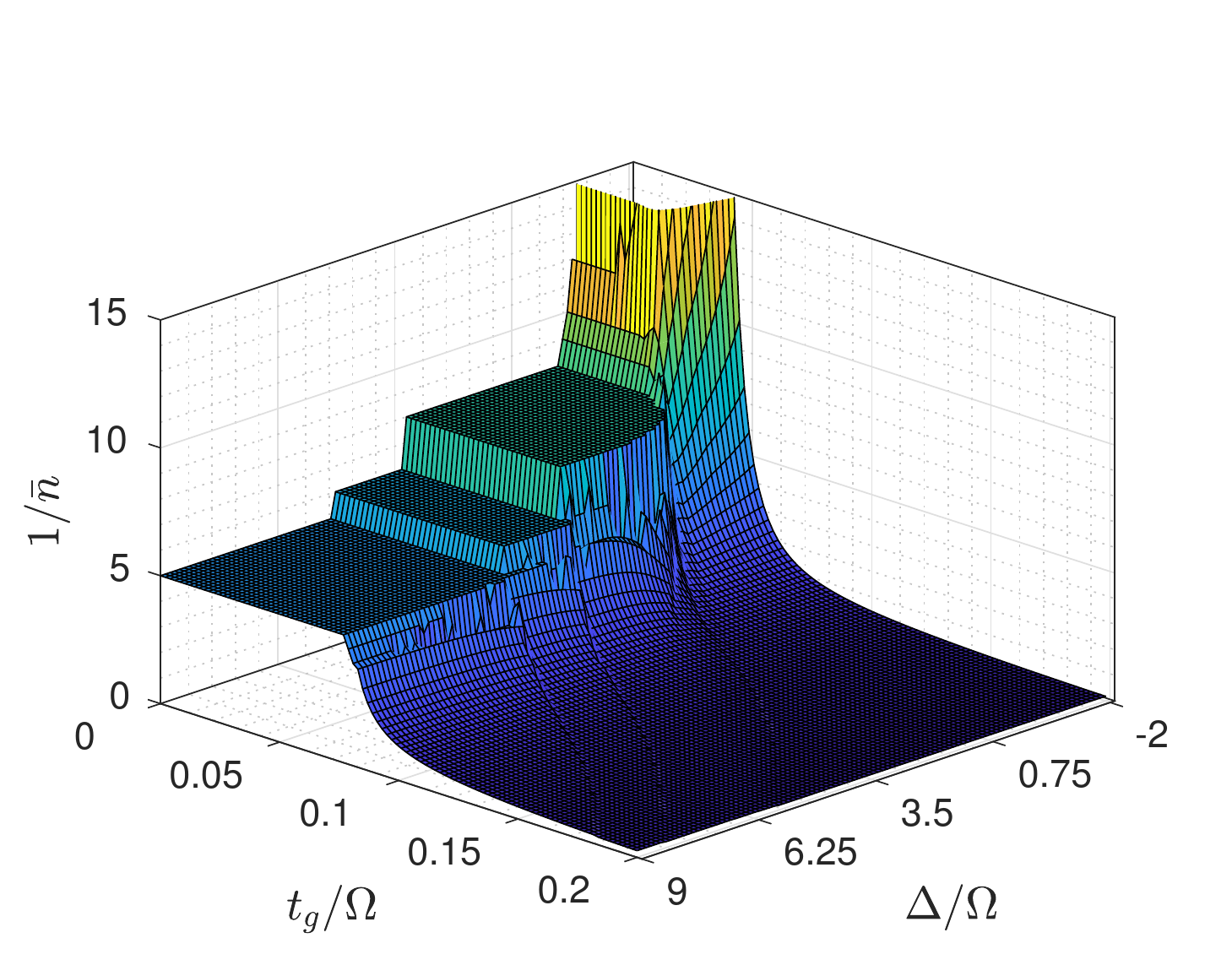}}
    \put(20,160){(a)}
    \end{picture}  }\\[-2.5ex]
\begin{tabular}{cc}
\subfloat{\begin{picture}(120,105)
        \put(0,0){\includegraphics[width=.25\textwidth,trim = {0 0 8 35}, clip]{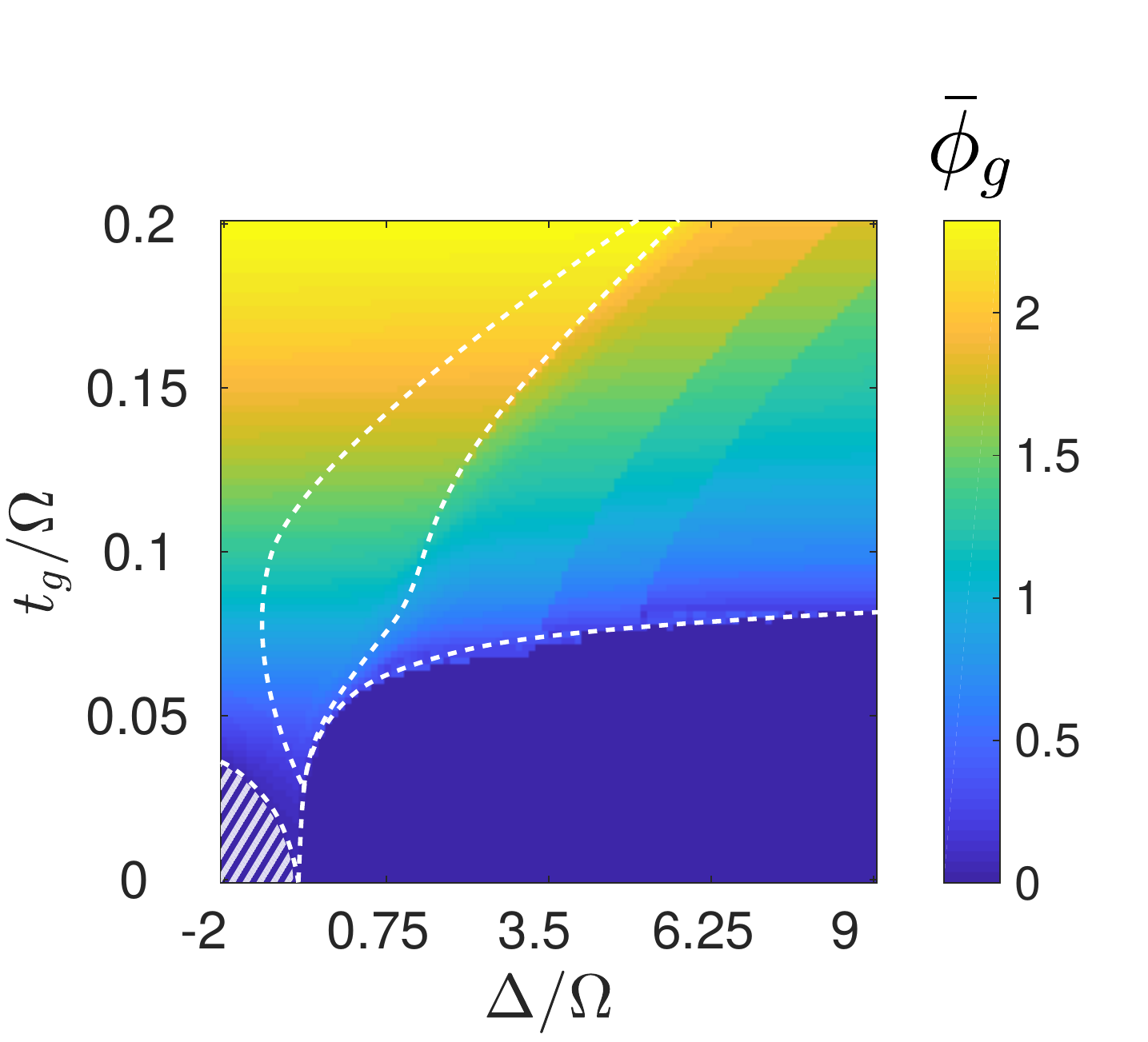}}
        \put(0,100){(b)}        
        \end{picture}}&
\subfloat{\begin{picture}(120,105)
        \put(0,0){\includegraphics[width=.25\textwidth,trim = {0 0 8 35}, clip]{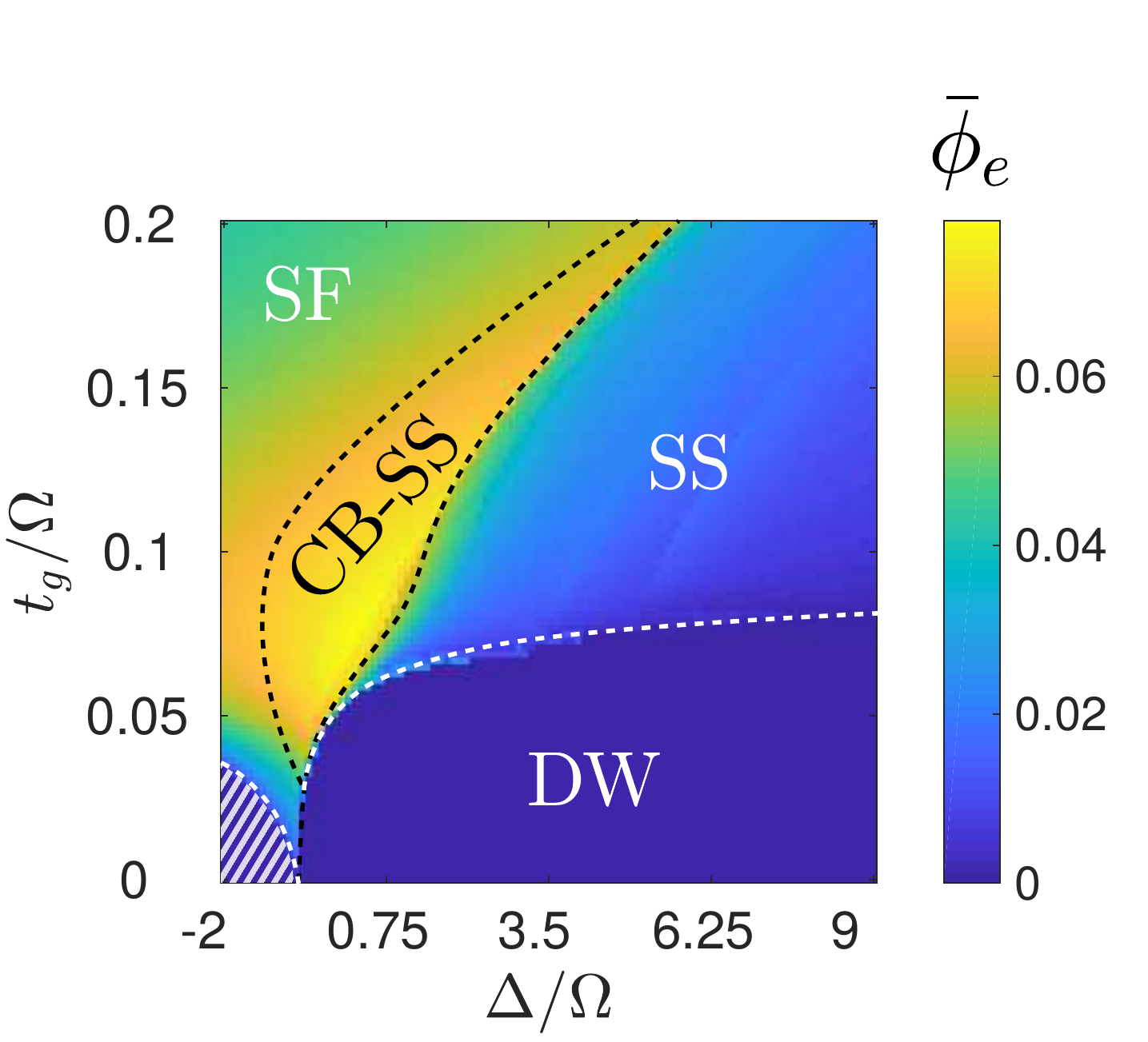}}
        \put(0,100){(c)}        
        \end{picture}}\\[-3.5ex]
\subfloat{\begin{picture}(120,105)
        \put(0,0){\includegraphics[width=.25\textwidth,trim = {0 0 8 40}, clip]{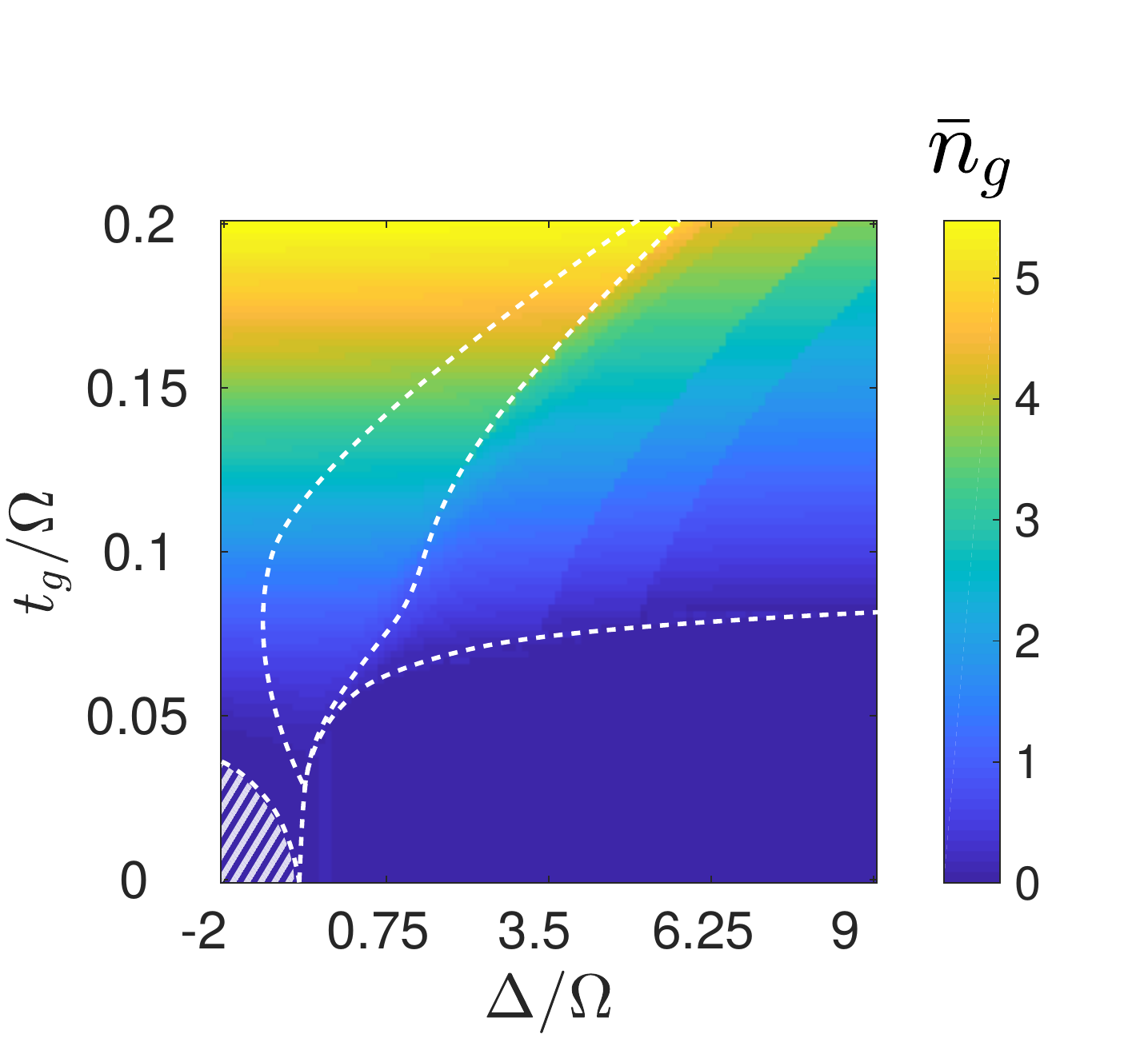}}
        \put(0,100){(d)}        
        \end{picture}}&
\subfloat{\begin{picture}(120,105)
        \put(0,0){\includegraphics[width=.25\textwidth,trim = {0 0 8 40}, clip]{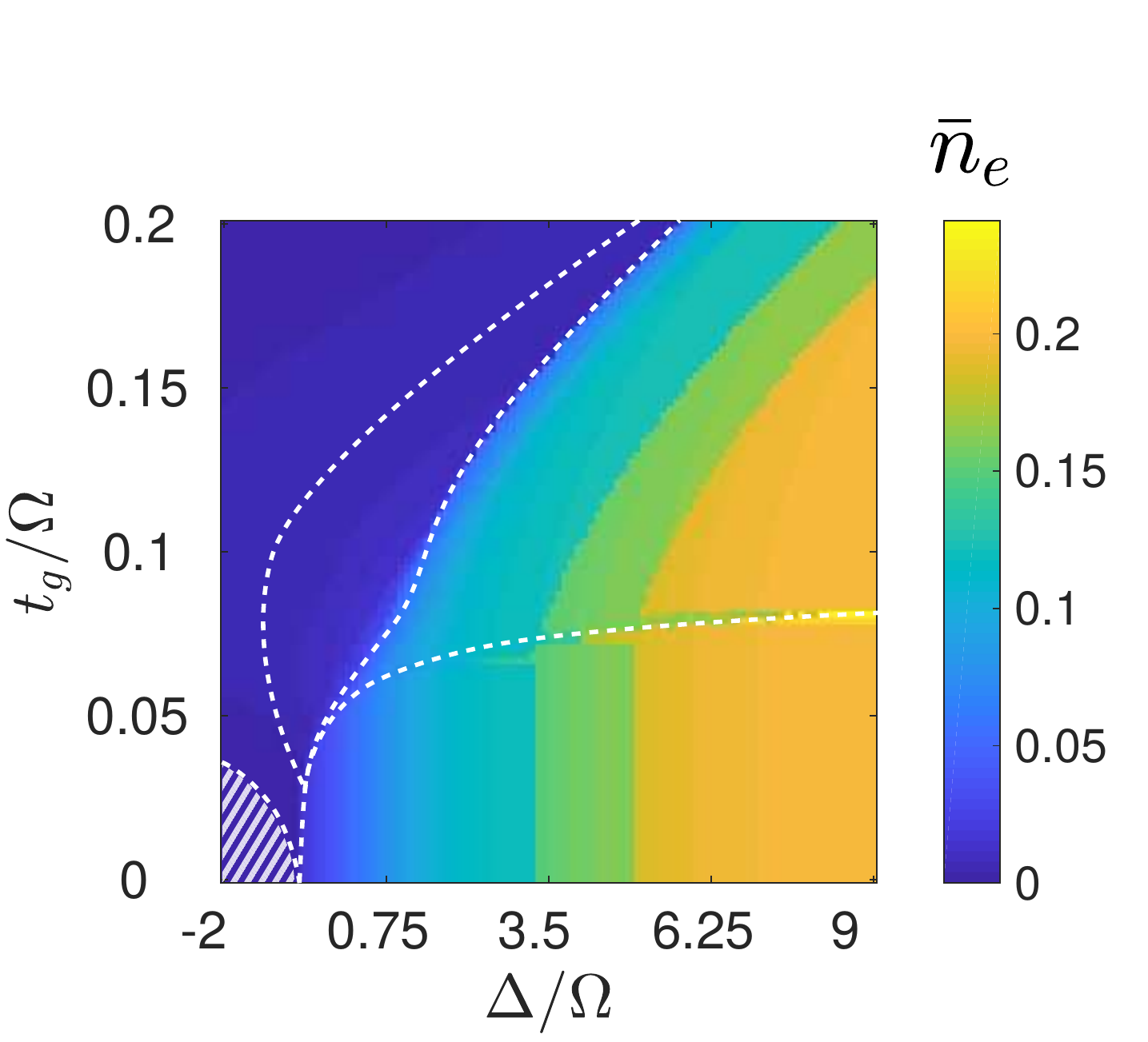}}
        \put(0,100){(e)}        
        \end{picture}}\\
\end{tabular}
\caption{Gutzwiller ground-state phase diagram: (a),the inverse density $1/\bar{n}$ of the many-body ground-state; (b)-(e), mean expectation values of the observables $\bar{\phi}_{\nu}$ and $\bar{n}_{\nu}$ for varying detuning $\Delta/\Omega$ and ground state hopping $t_g/\Omega$ and the drawn white phase boundaries act as rough guides to the eye. The fixed parameters for the calculations are $U_{g} = 0.1, U_{ge} = 5$, $U_{e} = 100$, $V = 10000$, $\mu = 0.25$,  $t_e = 0$ with $\Omega$ as scale.} \label{GutzwillerDiagram}
\end{figure}
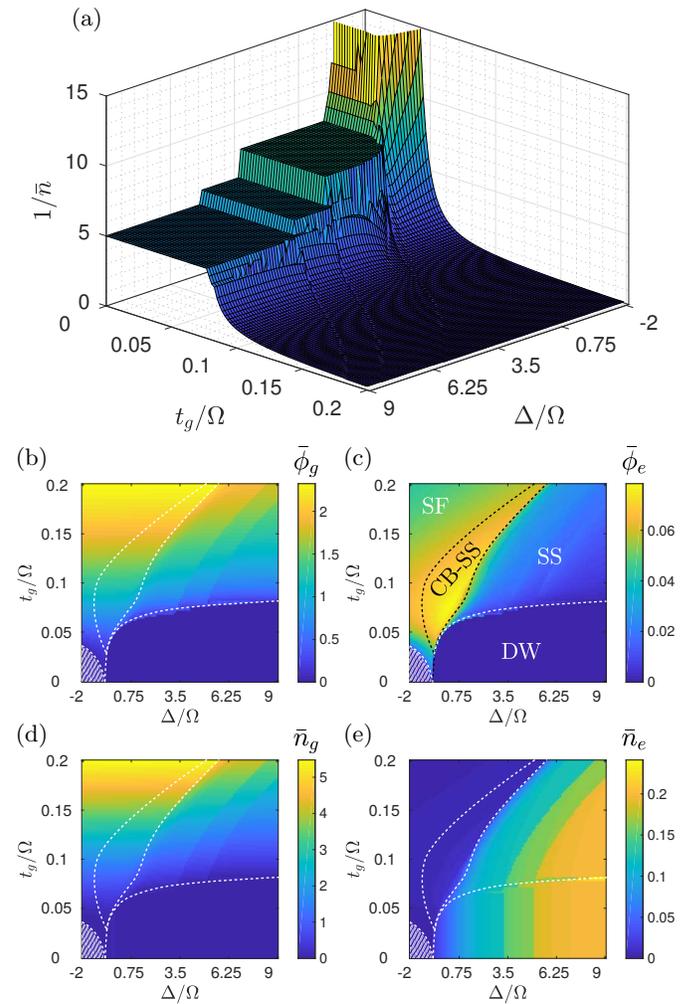
\begin{figure*}
\centering
  \includegraphics[width=0.99\textwidth,trim = {0 0 0 2}, clip]{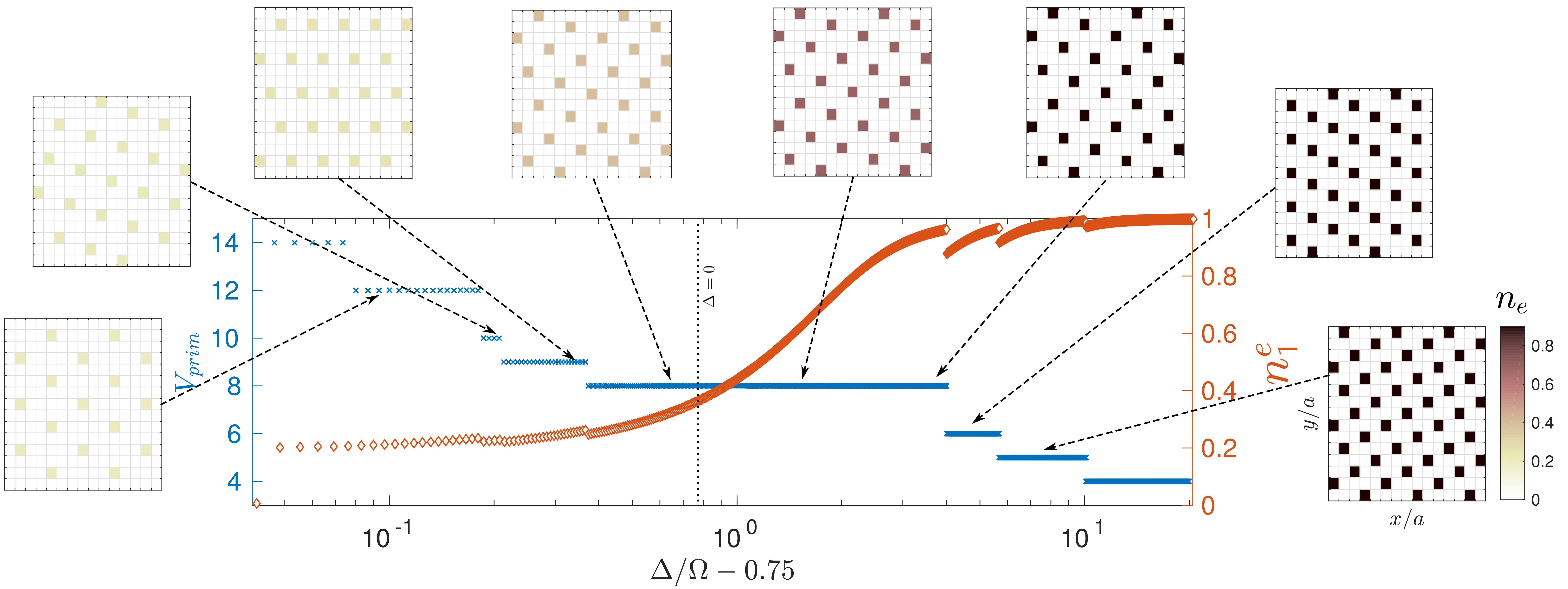}
  \caption{Devil's staircase of the primitive cell size $V_{prim}$ (blue crosses) and the Rydberg fraction on the occupied site $n^e_1$ (orange diamonds) in the "frozen" limit ($t_g/\Omega = 0$). Additionally, the real-space distributions of Rydberg excitations of various representative quantum phases are displayed in $x$- and $y$-direction in units of the lattice spacing $a$.} \label{Devils}
\end{figure*} 
In total, we consider 50 different structures of increasing primitive cell size. By varying the detuning $\Delta/\Omega$ and the ground state hopping rate $t_g/\Omega$ with all other parameters of the Hamiltonian $\hat{H}$ fixed, we obtain a rich phase diagram from the Gutzwiller calculations. We plot the inverse total density of the system $1/\bar{n}$ with $\bar{n} = \sum_{i,\nu} \langle \hat{n}^\nu_i \rangle/N $ in Fig. \ref{GutzwillerDiagram}(a). Combined with the mean condensate order parameter $\bar{\phi}_{\nu} =  \sum_i | \langle  \hat{b}^\nu_i \rangle |/ N $ and the mean occupation number $\bar{n}_{\nu} = \sum_i \langle \hat{n}^\nu_i \rangle / N$ depicted in Fig. \ref{GutzwillerDiagram}(b)-(e), we are able to identify different regimes of the phase diagram. In the "frozen" limit case, i.e. for vanishing hopping ($t_g/\Omega = 0$), the condensate order parameter of both states vanishes. Within this limit, variation of the detuning leads to a staircase-like modulation in particle density. High positive detuning favors a denser structure and a high Rydberg fraction $n_e$. Reducing the detuning results in narrow plateaus, terminating in a transition to the empty vacuum state. This transition can be traced back to the Hartree approximation, used to decouple the Hamiltonian \cite{Devil3}. Within the approximation, the relation between the chemical potential $\mu$, the detuning $\Delta$ and the Rabi frequency $\Omega$ at the transition reads
\begin{equation}
\mu = \frac{\Delta + \sqrt{\Omega^2 + \Delta^2}}{2}.
\end{equation}
For the chosen chemical potential $\mu/\Omega = 0.25$, the detuning at which the vacuum transition occurs is $\Delta/\Omega = -0.75$ and perfectly matches the numerical prediction. Approaching this transition from $\Delta/\Omega > -0.75$ leads to the so-called devil's staircase of lattice fillings \cite{Devil1,Devil2}. Decreasing the detuning does not increase the volume of the primitive cell gradually but rather in steps. We plot the staircase of the primitive cell size $V_{prim}$ and the Rydberg fraction $n^e_1$ on the occupied sites with higher resolution and a broader range of the detuning in Fig. \ref{Devils}. The steps become narrower as the transition to the vacuum state is approached. By including primitive cells of larger size, even narrower steps near the transition would be observed. This effect caused by the variation of the detuning is easily understandable from the Hamiltonian: Due to the RWA, the detuning acts as a chemical potential for Rydberg excitations and competes with the van-der-Waals interaction, causing dense structures with high Rydberg fractions for high positive detuning and sparse distribution of the Rydberg excitations for low or negative detuning. The drop of the Rydberg density through the reduction of the detuning can be achieved in two ways: By increasing the distance between nearby Rydberg excitations or by decreasing the on-site Rydberg fraction, the system is less subject to the long-range repulsion, as shown in Fig. \ref{Devils}.\\
Beyond the "frozen" limit, at finite hopping $t_g/\Omega > 0$, we find a variety of different quantum phases (see Fig. \ref{GutzwillerDiagram}(c)). The fluctuations induced by the hopping processes cause the narrow plateaus of quantum phases with low density near the transition to vanish, while the previously obtained staircase-like distribution of quantum phases further away from the vacuum transition remains unaffected. At a detuning-dependent critical hopping rate (for positive detuning around $t_g/\Omega \approx 0.6$), ground and excited state condensation emerge in the system. Since the system is completely localized below this critical hopping rate and at positive detunings, we can describe the quantum phases within this regime by particle density waves (DW). It is characterized by $\bar{\phi}_{\nu} = \phi_i^{\nu} = 0$, meaning vanishing local condensate order parameters. Above the critical hopping rate the particles start to delocalize. Furthermore, the staircase-like distribution of crystalline structures is shifted to larger values of the detuning. In this regime, both the long-range order and the finite condensate order parameter indicate a variety of supersolid quantum phases (SS), characterized by $\bar{\phi}_{\nu} \neq \phi_i^{\nu}$, indicating an inhomogeneous distribution of local condensate order parameters.\\
Around ($t_g/\Omega \approx 0.075,\Delta/\Omega \approx 0$) lies a regime in which small unit cell states with high mean condensate order parameter of the Rydberg state $\bar{\phi}_{e}$ exist. We find that the system favors a checkerboard-distribution of the Rydberg excitations. Due to the additional presence of a finite condensate order parameter, we conclude the presence of a checkerboard-supersolid quantum phase (CB-SS) in this regime. Sufficiently high hopping rates cause the system to become homogeneous and shift the phase boundary between supersolid and superfluid regime to higher positive detunings, rendering Rydberg excitations at any low or negative detunings unfavorable. Therefore, the system possesses a strongly reduced Rydberg fraction $n_e$, which leads to the loss of crystalline structures and hereby the homogenization of the system. The finite local ground-state condensate order parameter and the homogeneity imply a superfluid quantum phase in this regime (SF), characterized by $\bar{\phi}_{\nu} = \phi_i^{\nu} \neq 0$, meaning a homogeneous distribution of non-zero local condensate order parameters. The striped area in the lower right corner of the phase diagrams represents the vacuum state regime.\\
Overall, these results well match the results previously obtained with real-space bosonic dynamical mean-field theory (RB-DMFT) \cite{RydbergAndreas}.

\section{Decay-dephasing-induced steady states} \label{sec:IV}
We proceed to dynamical time evolution simulations within the GA. The inclusion of non-unitary channels in open quantum systems can be described with the master equation in Lindblad form (LB), a first-order differential equation
\begin{equation}
\frac{d\hat{\rho}(t)}{dt} = - \frac{i}{\hbar} [\hat{H},\hat{\rho}(t)] + \sum_k\Big(\hat{L}_k \hat{\rho}(t) \hat{L}_k^{\dag} - \frac{1}{2} \{\hat{L}_k^{\dag}\hat{L}_k,\hat{\rho}(t) \} \Big). \label{eq:Lindblad}
\end{equation}
The Lindblad operator $\hat{L}_k$ describes the $k^{th}$ non-unitary channel, which couples the system to its environment. In our case $\hat{H}$ corresponds to the full Hamiltonian \eqref{eq:Hamiltfull} and the time-dependent density matrix $\hat{\rho}(t)$ is that of the whole system. Motivated by the previous calculations of the non-dissipative Gutzwiller ground-states, we solve the Lindblad equation \eqref{eq:Lindblad} approximately by decoupling the lattice sites \cite{DissipativeRydbergAtoms}. Applying the Gutzwiller ansatz to the density matrix $\hat{\rho}(t) = \bigotimes_i \hat{\rho}_i(t)$, initially consisting of the single-site terms $\hat{\rho}_i(0) = |\Psi(0)\rangle_i \langle \Psi(0)|_i$, and using the single-site Hamiltonian $\hat{H}_i$, a decoupled Lindblad equation can be derived
\begin{equation}
\begin{split}
\frac{d\hat{\rho}_i(t)}{dt} &= - \frac{i}{\hbar} [\hat{H}_i,\hat{\rho}_i(t)]\\
 &+ \sum_k\Big(\hat{L}_{i,k} \hat{\rho}_i(t) \hat{L}_{i,k}^{\dag} - \frac{1}{2}  \{\hat{L}_{i,k}^{\dag}\hat{L}_{i,k},\hat{\rho}_i(t) \} \Big). 
\end{split} \label{eq:LindbladDecoupled}
\end{equation}\noindent
We consider two non-unitary channels $\hat{L}_{i,k}$, the first one being the decay of the Rydberg state to the ground state due to the finite lifetime of the excited state. The photon emitted during the decay leaves the system and hereby entangles it to the environment. The corresponding Lindblad operator
\begin{equation}
\hat{L}_{i,dec} = \sqrt{\Gamma} (\hat{b}^g_i)^\dag \hat{b}^e_i
\end{equation}
annihilates a particle in the Rydberg state and creates one in the ground state with rate $\Gamma$. This corresponds to deexcitation through spontaneous emission and is the first non-unitary contribution.\\
Note that a decaying atom acquires a momentum kick through emission of a photon and hereby gains energy given by the recoil energy $E_R$. Since we consider only the lowest Bloch band, the optical lattice has to be sufficiently deep in order to guarantee that the band gap is bigger than the energy gained through the momentum kick, thus preventing possible transitions of the atom to the second band. Optical lattice depths above $V_0 = 5 E_R$ ensure the required band gap \cite{BlochBands}.\\
Two-photon excitation schemes are commonly used to reach a Rydberg state with isotropic interaction in this field. Despite the considered incoherent effects, these schemes may cause additional complications such as atom loss from the optical trap and population of degenerate ground states not considered in the model. While leakage from the considered internal degrees of freedom can be kept under control by a depumping and repumping scheme \cite{RepumpI,RepumpII,RepumpIII,RepumpIV}, control of the trap loss is a more demanding task and beyond the scope of the present paper. A possible method to get rid of both effects is to switch to a single-photon excitation scheme instead. Even though the adressed Rydberg state possesses a non-zero angular momentum quantum number the interaction can be made isotropic in 2D by application of an external magnetic field \cite{IsotropyI,IsotropyII}.\\
Additionally, a collective dephasing of particles in the Rydberg state has been observed in several experiments \cite{DephasingI,DephasingII,Bistability}. There are several sources for this effect, one being the linewidth of the excitation laser. Due to the finite linewidth different Rydberg states are coherently addressed, which quickly dephase before and after going back to the ground state \cite{LamourPhD}. Theoretically, the dephasing can be explained by means of the Hamiltonian. Since solely the interaction of one Rydberg state is included in the description of the system, the interaction with the non-included Rydberg states causes an effective dephasing that scales, similar to an interaction, with the Rydberg density of the system. The corresponding Lindblad operator best modelling this effect reads
\begin{equation}
\hat{L}_{i,deph} = \sqrt{\kappa} \hat{n}^e_i,
\end{equation}
where $\kappa$ is the dephasing rate. We do not consider here the avalanche-like atom- and condensate-loss caused by the broadening, which has been observed and discussed in various works \cite{AvalancheII,FrozenGasI,FrozenGasII}. The branching of the Rydberg state to contaminant states can be limited in a cryogenic environment \cite{Cryogenic}.\\
Solving the decoupled Lindblad equation \eqref{eq:LindbladDecoupled} can be approached in several ways. By setting $d\hat{\rho}(t)/dt = 0$ one can obtain an implicit function containing the density matrix of the steady states of the system. Subsequent calculations of the mean condensate order parameters $\bar{\phi}^{\nu}_{ss} = \lim_{t \rightarrow \infty} \sum_{i = 1}^N \text{tr}(\hat{\rho}(t) \hat{b}^\nu_i)/N$ indicate that the steady state is always fully localized and does not possess any condensate for finite non-unitary rates ($\bar{\phi}^{\nu}_{ss} = 0$). However, in doing so, we lose valuable information about the dynamical evolution into this steady-state. Since we primarily focus on experimentally states that are long-lived, meaning nearly constant behavior of all considered observables, a time evolution simulation is necessary.\\
Using the 4$^{\text{th}}$ order Runge-Kutta method, we obtain the discretized dynamics of the density matrices
\begin{equation*}
\begin{split}
\hat{\rho}_{t+\Delta t} &= \hat{\rho}_t + \Delta t \sum_{i=1}^4 b_i k_i\\
k_i &= f(t + c_i\Delta t, \hat{\rho}_t + \Delta t \sum_{j=1}^{4} a_{ij} k_j)
\end{split}
\end{equation*}
with $f(\cdot,\cdot) = d\hat{\rho}(t)/dt$ given by the right-hand side of the Lindblad equation \eqref{eq:LindbladDecoupled}. Note that time is discretized into timesteps $t_i$ of length $\Delta t$. Furthermore, we choose the Runge-Kutta coefficients $a_{ij}$ and $b_i$ in such a way that the quadratic first integrals of the density matrices are conserved ($I(\hat{\rho}_{t_i}) = I(\hat{\rho}_{t_j})$, where $I(\hat{\rho}_t) = \hat{\rho}^T_t C \hat{\rho}_t$ with an arbitrary symmetric matrix $C$), and the method hereby becomes symplectic. This ensures the numerical conservation of all physically conserved quantities, such as the total particle number, throughout the simulation \cite{Symplectic1,Symplectic2,Schlick}. In the 4$^{\text{th}}$ order Runge-Kutta method, this is equivalent to the relation $(b_i b_j - b_i a_{ij} - b_j a_{ji}) = 0$.\\
We use the Gutzwiller ground-states $|\Psi_i \rangle$ calculated for the phase diagram in Fig. \ref{GutzwillerDiagram} as initial pure-state density matrices $\hat{\rho}_{i}(0) = |\Psi\rangle_i \langle \Psi|_i$ of each site, perform the time evolution simulations and analyze the dynamics in terms of $\phi^\nu_{i}(t) = \text{tr}(\hat{\rho}_{i}(t) \hat{b}^\nu_i)$ and $n^\nu_{i}(t) = \text{tr}(\hat{\rho}_{i}(t) \hat{n}^\nu_i)$.\\
In what follows, we focus on two specific supersolid phases: A checkerboard-structured supersolid and a supersolid with a more complex crystalline structure. We set the Rabi frequency to $\Omega = 1$MHz and use the natural units, hence $\hbar = 1$.

\subsection{Decay-dephasing-quench time evolution} \label{ch:decay-dephasing}
\begin{figure}[h]
    \centering
    \includegraphics[scale=0.52, trim = {0.5cm 0.18cm 0.5cm 0cm}, clip]{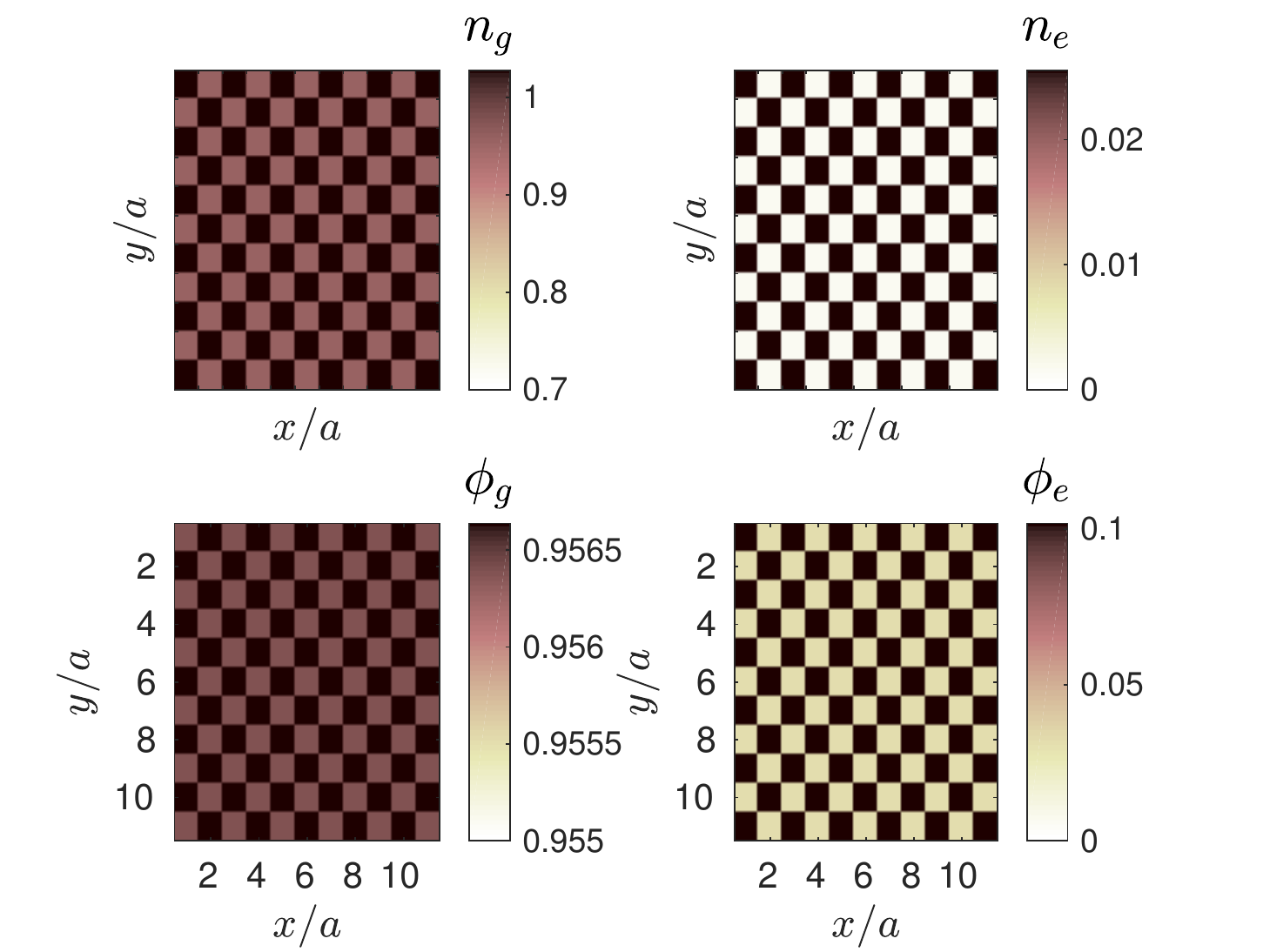}
    \caption{Initial CB-SS quantum phase with average filling $\bar{n} = 1$. The primitive cell of the system consists of two sites. The parameters for the Gutzwiller calculations are $U_{g} = 0.1, U_{ge} = 5$, $U_{e} = 100$, $V = 10000$, $\mu = 0.1$, $t_g = 0.0414$, $t_e = 0$, $\Delta = 0.069$ with $\Omega$ as scale.} \label{fig:initial-checkerboard}
\end{figure}

We choose as an initial state a CB-SS with average filling of $\bar{{n}} = 1$, as depicted in Fig. \ref{fig:initial-checkerboard}. We vary the rates of both non-unitary processes, the spontaneous emission rate $\Gamma/\Omega \in [0,5]$ and the dephasing rate $\kappa/\Omega \in [0, 20]$. Those parameter ranges only serve the purpose of identifying the influence of both non-unitary processes. Values close to experimentally relevant rates will be used later on.
The time evolution begins with the sudden enabling of the non-unitary channels, starting with the Gutzwiller ground-state of the CB-SS phase. Since the checkerboard-structure is the simplest crystalline one on the square lattice and we expect the non-unitary processes to destroy long-range order and thus reduce the crystalline order, this implies a transition into a homogeneous system. In order to understand whether the CB-SS can be preserved or is always unstable under the influence of those processes, we perform time evolution up to a final time $t_f = 100\mu s$ and determine the observables $\langle \hat{n}_i^\nu (t_f)\rangle$ and $\langle \hat{b}_i^\nu (t_f)\rangle$ with $\nu \in \{g,e\}$ on both sublattices $i \in \{1,2\}$.\\
We visualize whether the system is homogeneous or not by plotting the imbalance $\Delta n = |n_{1}(t_f) - n_{2}(t_f)|$ with $n_i = \sum_\nu \langle \hat{n}^\nu_i (t_f)\rangle$, which is the difference between the total occupation numbers per site of both sublattices, for various decay and dephasing rates. Plotted in Fig. \ref{fig:cbss-maps}(a) is the imbalance of the checkerboard. Obviously, high decay and dephasing rates lead to homogeneity, which confirms our expectations. For intermediate rates, the imbalance is enhanced and increases even further for lower rates, peaking around $(\Gamma/\Omega, \kappa/\Omega) = (0.5,6)$. 
\begin{figure}[t]
\centering
\begin{picture}(205,175)
\put(0,10){\includegraphics[width = 0.4\textwidth, trim = {0cm 0cm 0.7cm 1cm}, clip]{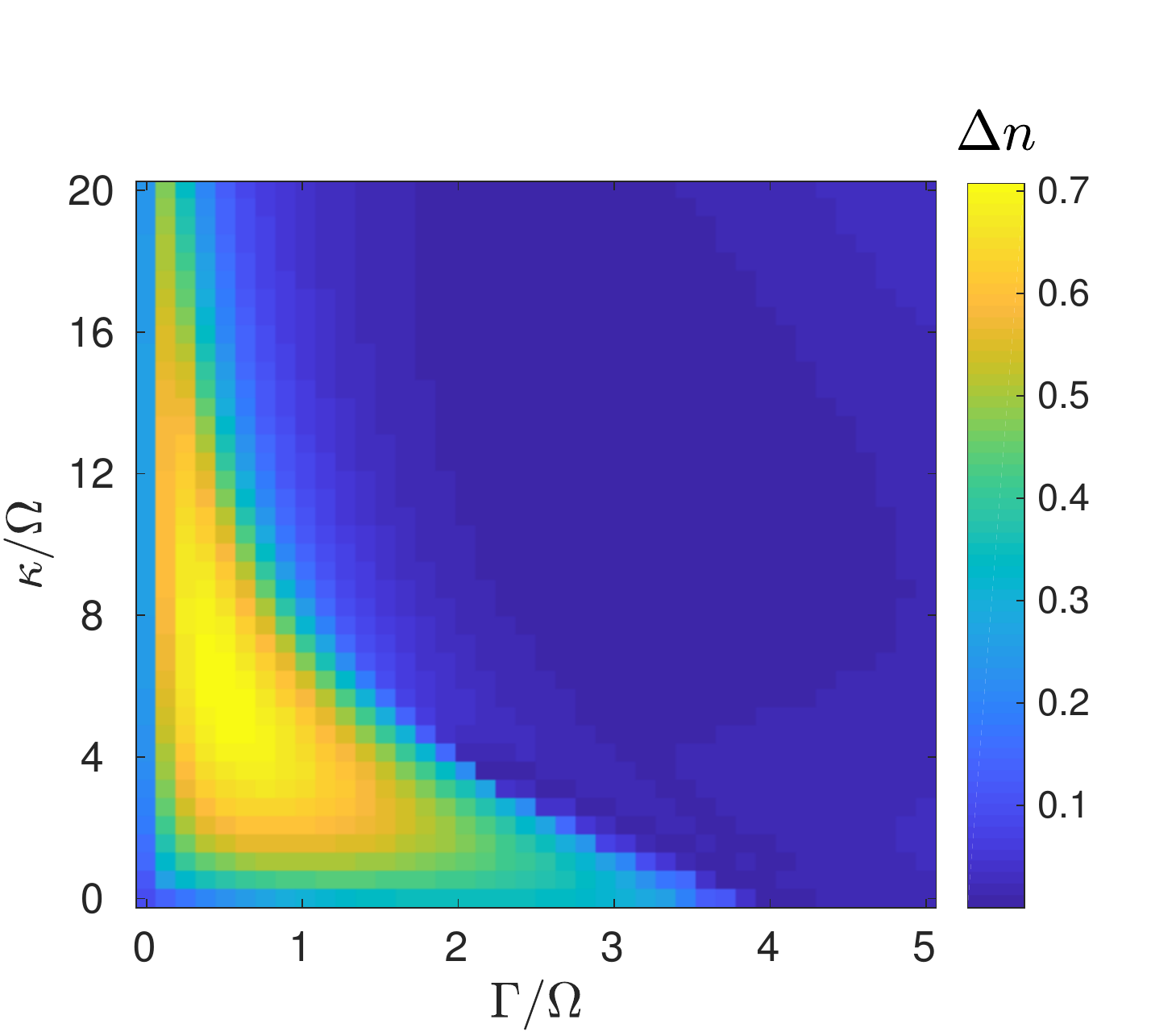}}
\put(5,175){(a)}
\end{picture}
\begin{picture}(205,165)
\put(0,0){\includegraphics[width = 0.4\textwidth, trim = {0cm 0cm 0.7cm 1cm}, clip]{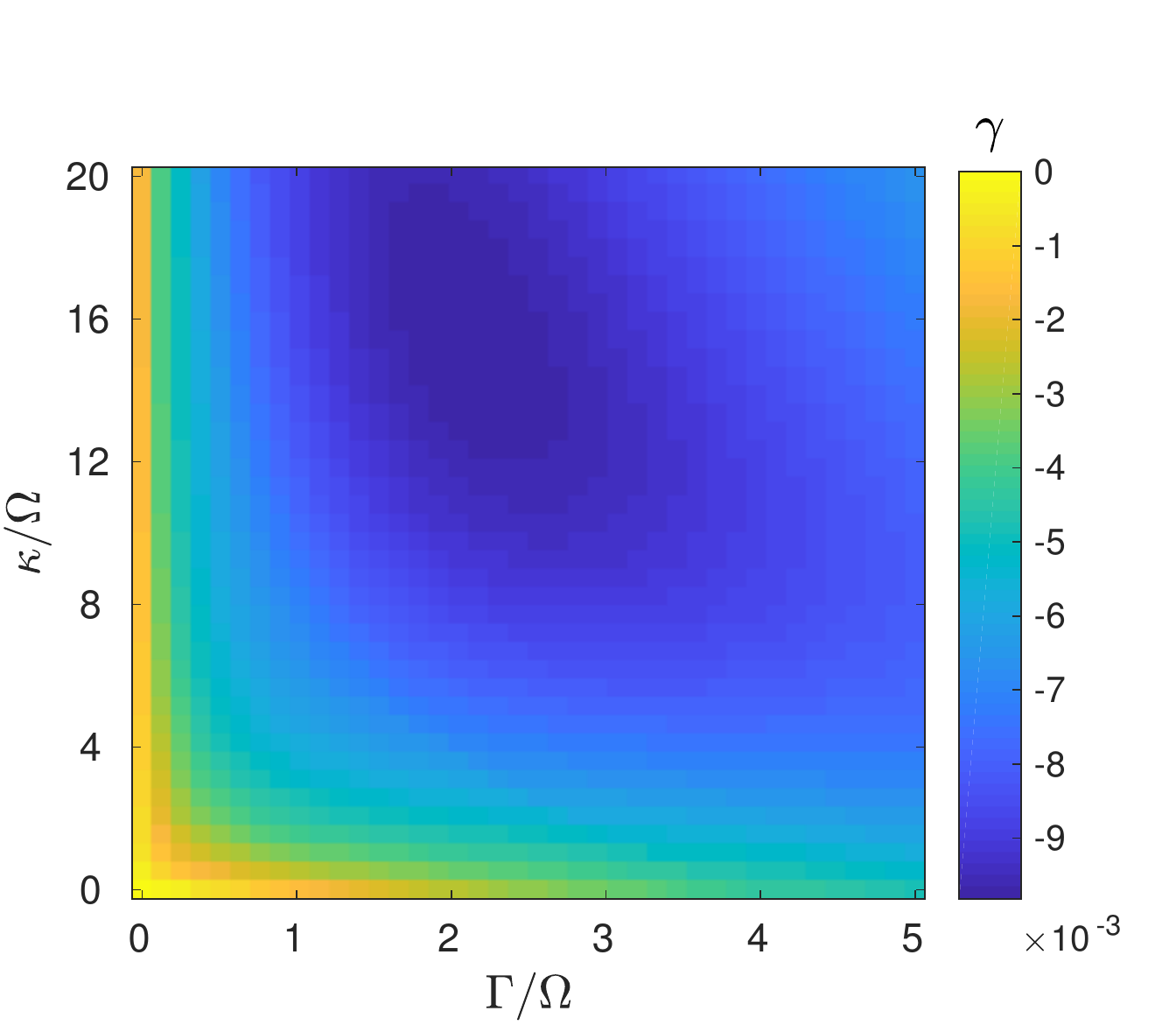}}
    \put(5,165){(b)}
    \end{picture}
    \caption{(a) imbalance $\Delta n(t_f)$ as a function of the decay and dephasing rates; (b) the exponential decay constant of the mean condensate order parameter of the ground state $\gamma$ as a function of the decay and dephasing rates ($t_f = 100 \mu s$).} \label{fig:cbss-maps}
\end{figure}
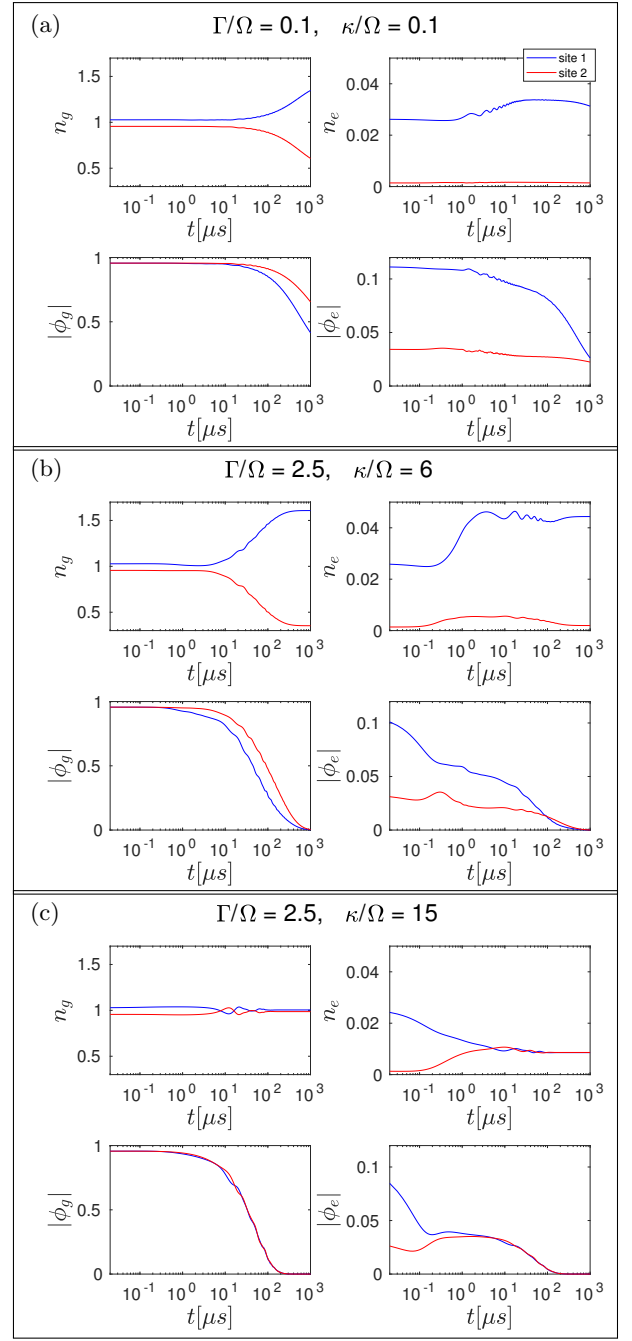
\begin{figure}[t]
\subfloat{
        \fbox{
        \begin{picture}(219,163)
        \put(0,0){\includegraphics[width=.45\textwidth, trim = {0.5cm 0.2cm 0.5cm 0.2cm}, clip]{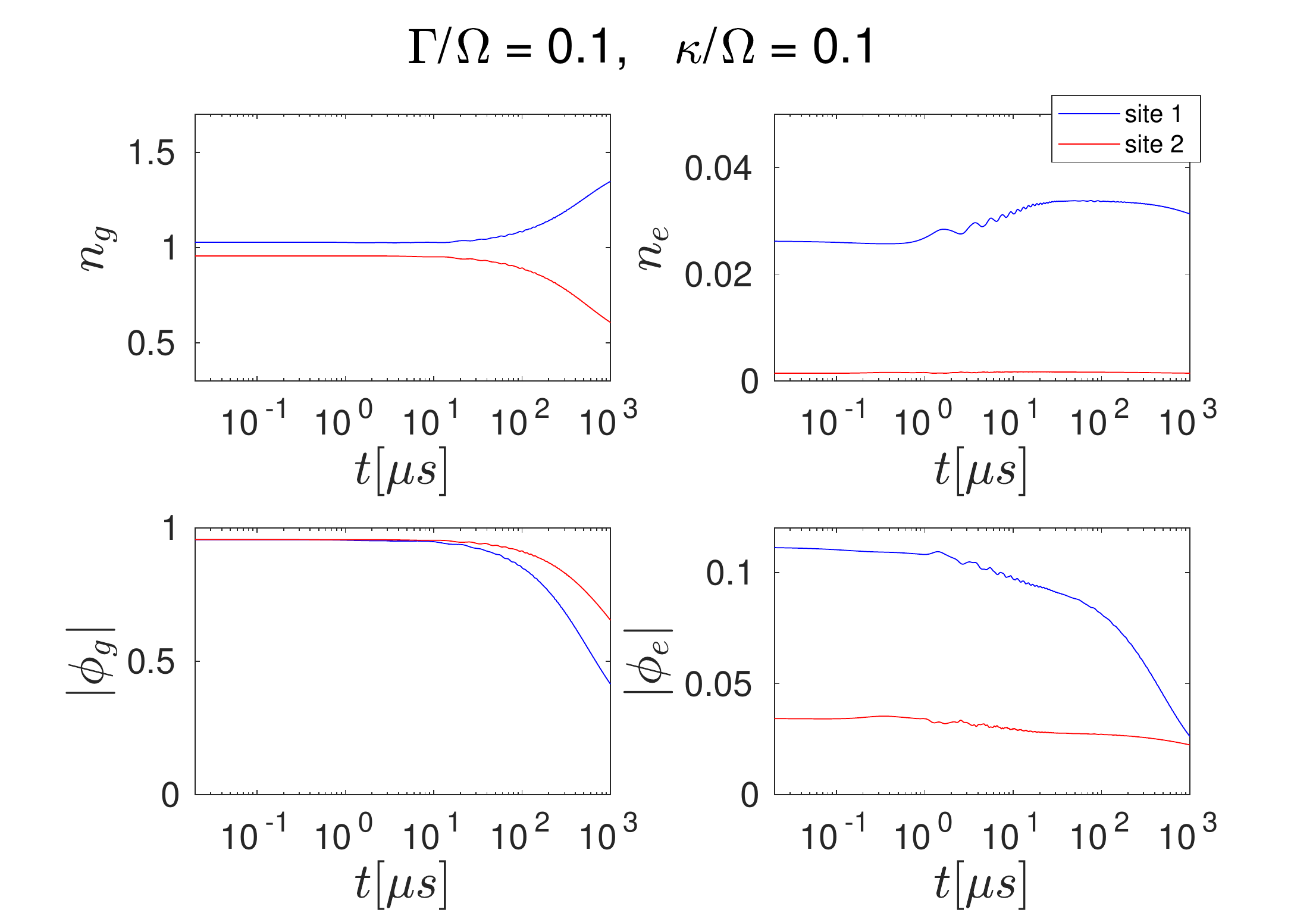}}
        \put(0,155){(a)} 
        \end{picture}}}\\ [-0.7ex]
\subfloat{
        \fbox{
        \begin{picture}(219,163)
        \put(0,0){\includegraphics[width=.45\textwidth, trim = {0.5cm 0.2cm 0.5cm 0.2cm}, clip]{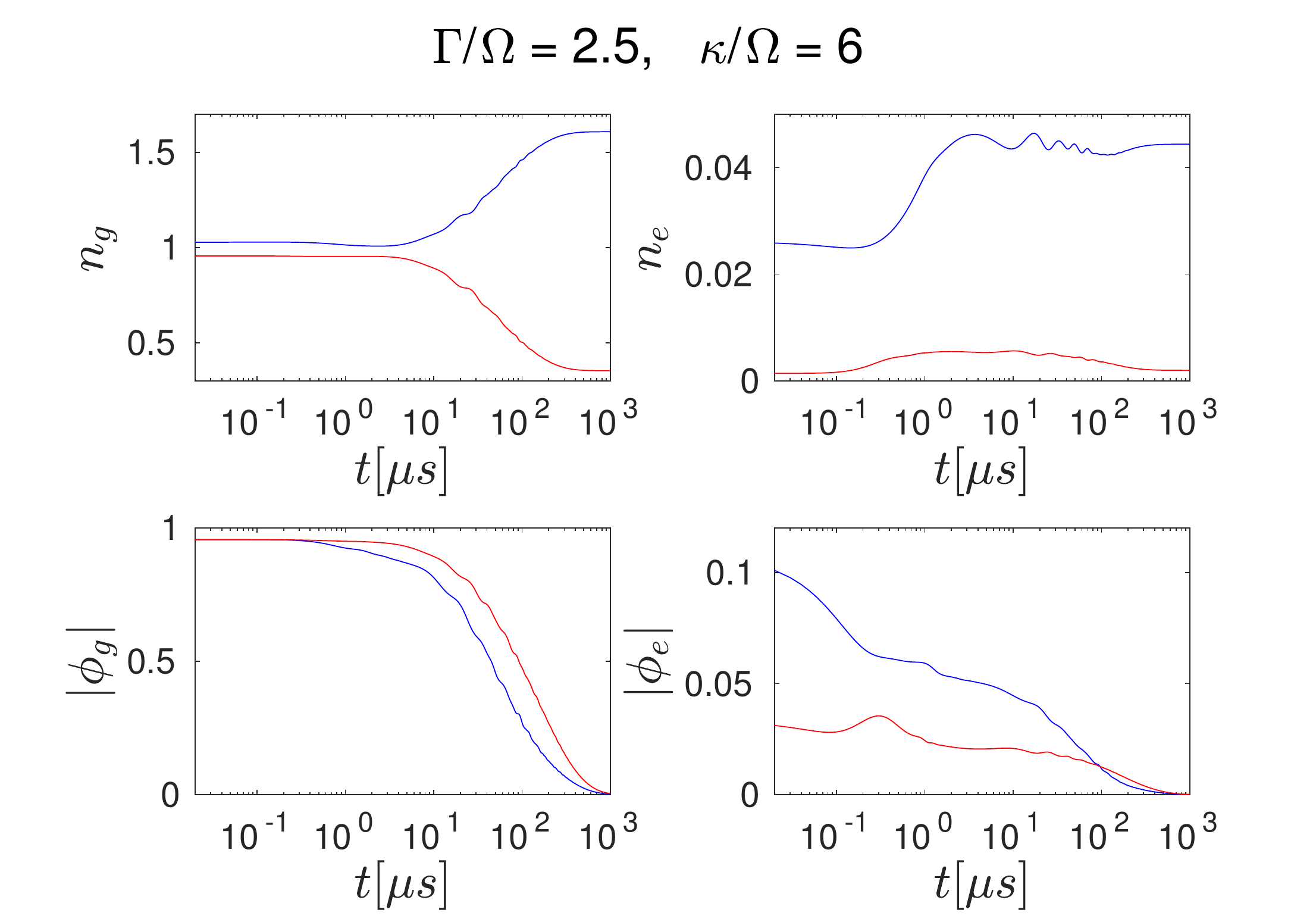}}
        \put(0,155){(b)}
        \end{picture}}}\\ [-0.7ex]
\subfloat{
        \fbox{
        \begin{picture}(219,163)
        \put(0,0){\includegraphics[width=.45\textwidth, trim = {0.5cm 0.2cm 0.5cm 0.2cm}, clip]{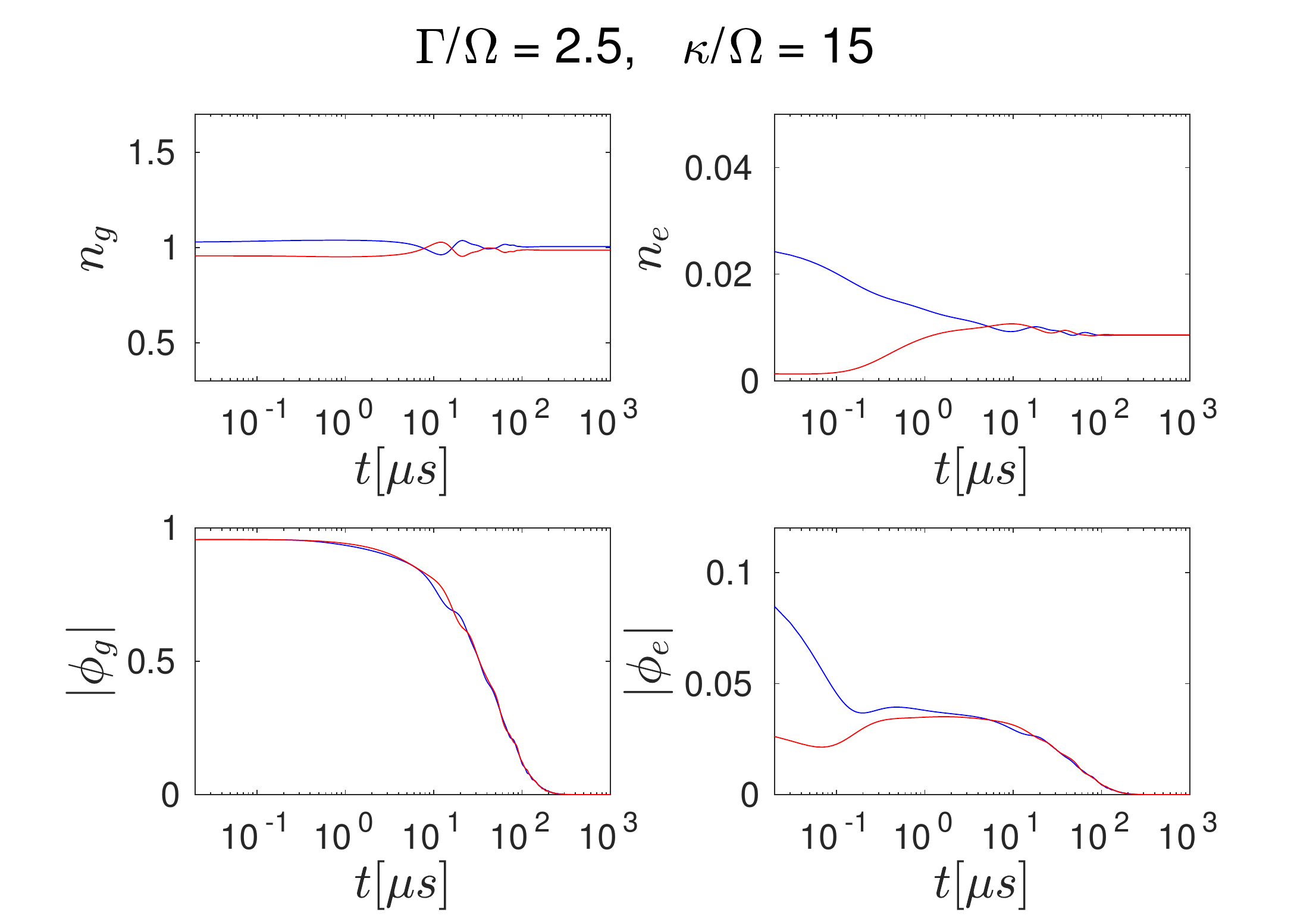}}
        \put(0,155){(c)}
        \end{picture}}}\\      
    \caption{Time evolution of relevant observables for an initial CB-SS state shown in Fig. \ref{fig:initial-checkerboard} for different decay and dephasing rates $\Gamma/\Omega$ and $\kappa/\Omega$. Low rates cause a slow decay of the condensate and increase the imbalance $\Delta n$ of the system, moderate rates enhance the speed with which the condensate drops, and high rates lead to homogeneity.} \label{fig:cbss-evolution}
\end{figure}
We perform an exponential fit $\bar{\phi}_g(t) = \phi_0 e^{\gamma t}$ of the mean condensate order parameter of the ground state with $\phi_0 = \bar{\phi_g}(t = 0)$, since we assume its behavior to be exponential for now. In Fig. \ref{fig:cbss-maps}(b), the exponential decay constant $\gamma$ is mapped versus the decay and the dephasing rate. A loss of condensate order parameter is observable for any non-unitary channel enabled. The exponential decay constant grows by increasing the decay and dephasing rates to intermediate values, but decreases for even higher rates. We are able to quantitatively discuss the dependence of the speed of the condensate loss to the decay and dephasing rates: Reminescent of the QZE, high rates would imply a strong and continuous measurement of the system, which effectively freezes the system. As a result, the exponential decay constant i.e. the speed of the condensate loss eventually becomes smaller for further increasing rates of the non-unitary processes. The exponential decay constant is maximal at about $(\Gamma/\Omega, \kappa/\Omega) = (2.5,15)$.\\
In order to better understand the effect of the decay and dephasing, we perform and depict the dynamics for a few representative decay and dephasing rate pairs ($\Gamma/\Omega, \kappa/\Omega$) at particular parameter values previously mentioned (see Fig. \ref{fig:cbss-evolution}). First off, we see in all cases the previously assumed exponential decay behavior of the ground state condensate order parameter $\phi_g$, confirming our expectations. We begin with the time evolution of the CB-SS quantum phase subject to spontaneous decay and dephasing with low rates (see Fig. \ref{fig:cbss-evolution}(a)). The difference in the occupation numbers e.g. the imbalance $\Delta n$ increases steadily, while all local condensate order parameters $\phi_g$ and $\phi_e$ slowly decay. Moving into the regime where the imbalance $\Delta n$ peaks, we observe a transition to a state where all considered observables tend to constant values, thus we consider it a long-lived state (see Fig. \ref{fig:cbss-evolution}(b)). Compared to the time evolution with lower non-unitary rates in Fig. \ref{fig:cbss-evolution}(a), an earlier and faster decay of all local condensate order parameters $\phi_g$ and $\phi_e$ at intermediate times can be seen. At later times the system lost all its condensate and hereby becomes a inhomogeneous DW quantum phase. We continue with the regime of high decay and dephasing rates (see Fig. \ref{fig:cbss-evolution}(c)). The non-unitary processes cause a drop of the Rydberg state population $n_e$ within a short time, which leads to a loss of crystalline structure. The ground state occupation number $n_g$ oscillates around its average due to the quench of the decay and dephasing rates and relaxes after approximately 150$\mu s$. Paired up with the loss of condensate, we observe the dynamical phase crossover to a homogeneous DW quantum phase.\\
We observe a decay of all local condensate order parameters induced by either non-unitary channel. Additionally, we observe that the spatial distribution of particles remains unaffected in the case of low rates of the non-unitary processes, while high rates cause a rapid drop of the Rydberg fraction $n_e$, ultimately leading to the loss of long-range order.\\
\par
\begin{figure}[t]
    \centering
    \includegraphics[scale=0.55, trim = {0.5cm 0.18cm 0.5cm 0cm}, clip]{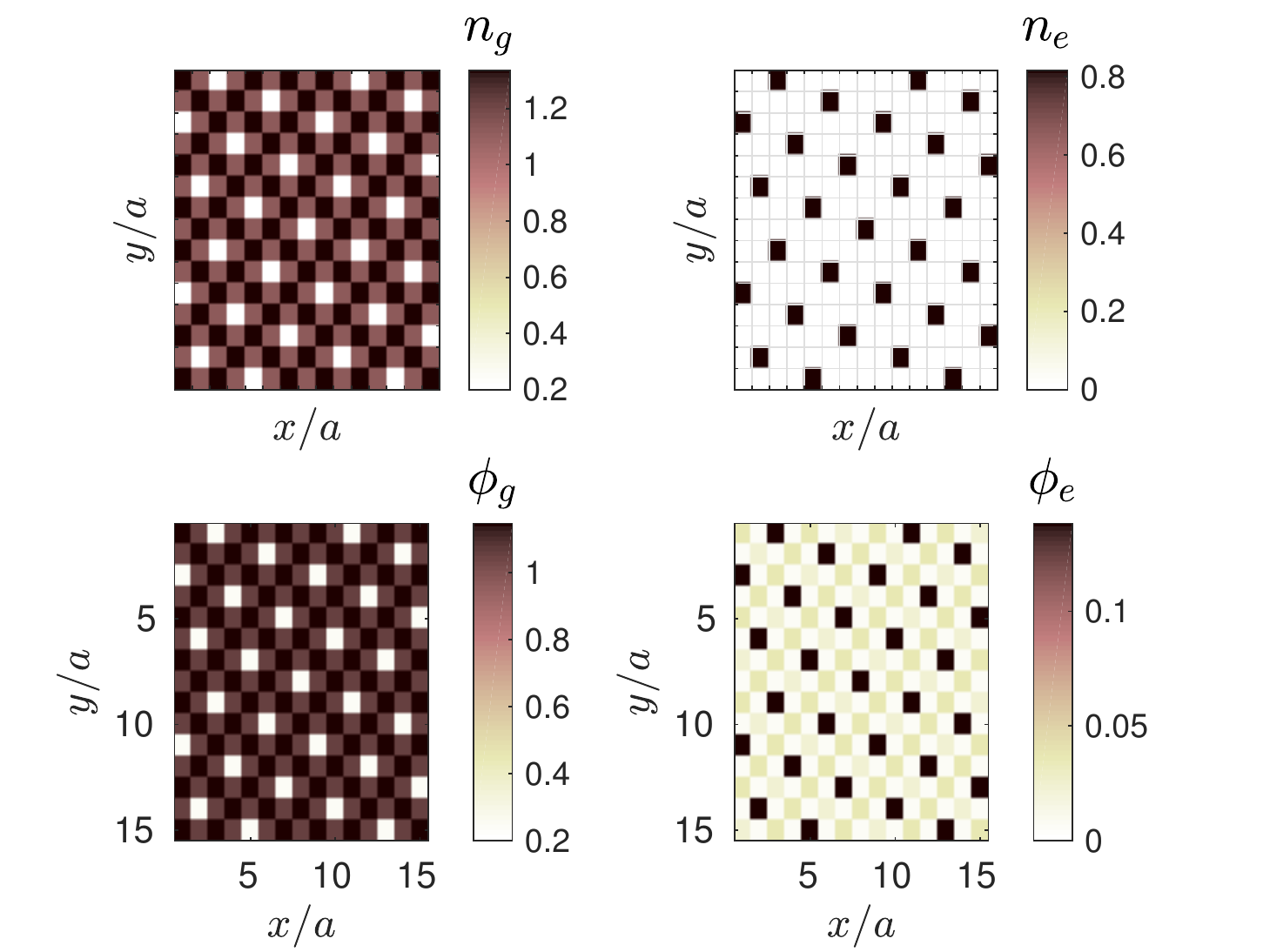}
    \caption{Initial SS quantum phase with average filling $\bar{n} \approx 1.2$. The primitive cell of the system consists of eight sites. The parameters for the Gutzwiller calculations are $U_{g} = 0.1, U_{ge} = 5$, $U_{e} = 100$, $V = 10000$, $\mu = 0.25$, $t_g = 0.099$, $t_e = 0$, $\Delta = 2.18$ with $\Omega$ as scale.} \label{fig:initial-supersolid}
\end{figure}

We now focus on one representative SS with a more complex structure, the geometry of which is described by four sublattices, as depicted in Fig. \ref{fig:initial-supersolid}. In contrast to the CB-SS, where the transition into a homogeneous system has been investigated through the calculation of the imbalance, here the best way to analyze the effects of the non-unitary channels is through the depiction of the dynamics.
\begin{figure}[t]
\captionsetup[subfigure]{labelformat=empty}
\begin{tabular}{ccc}
\subfloat[$t = 0 \mu s$]{
\includegraphics[width = .14\textwidth, trim = 10 10 0 0, clip]{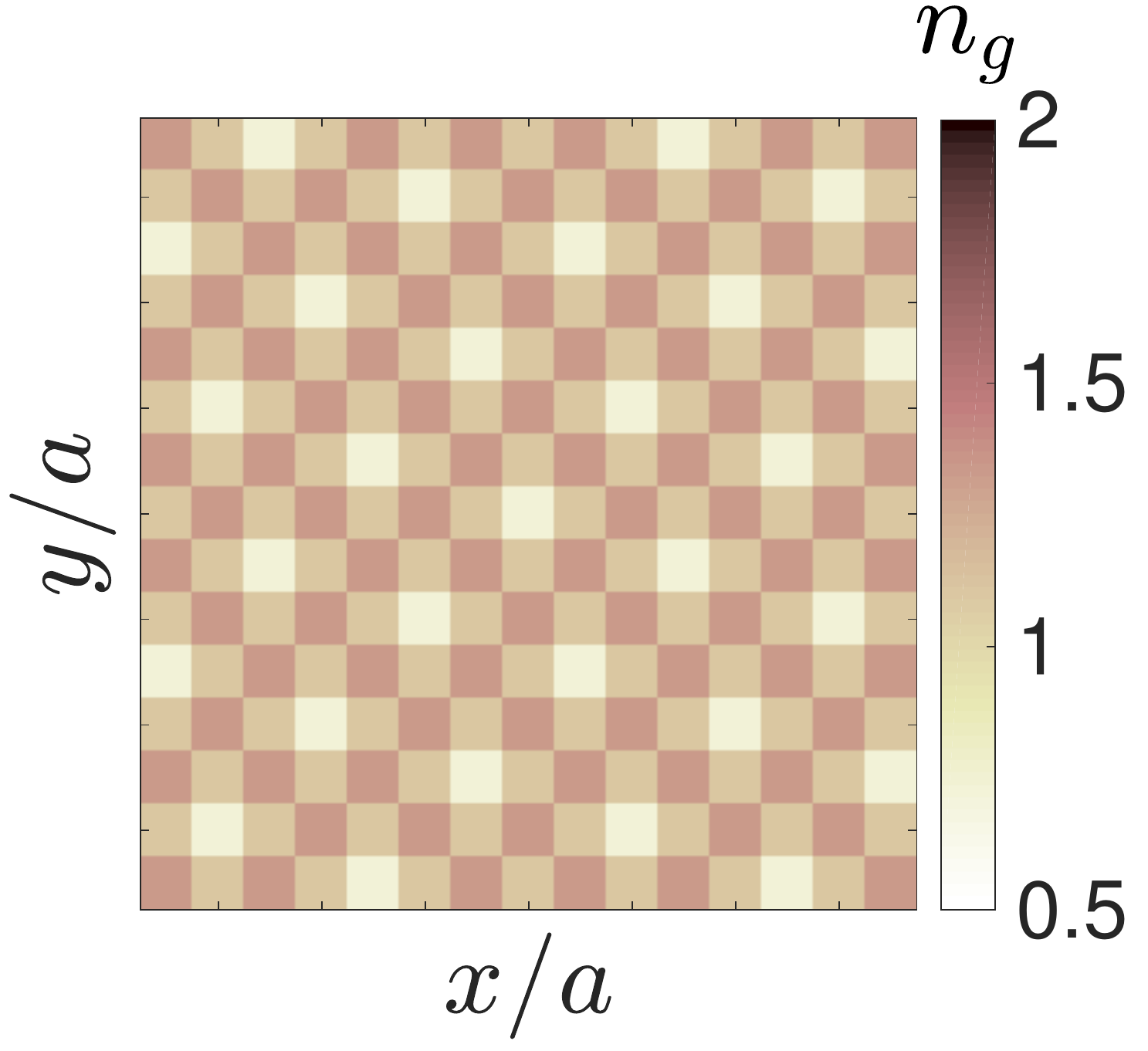}}&
\subfloat[$t = 20 \mu s$]{
\includegraphics[width = .14\textwidth, trim = 10 10 0 0, clip]{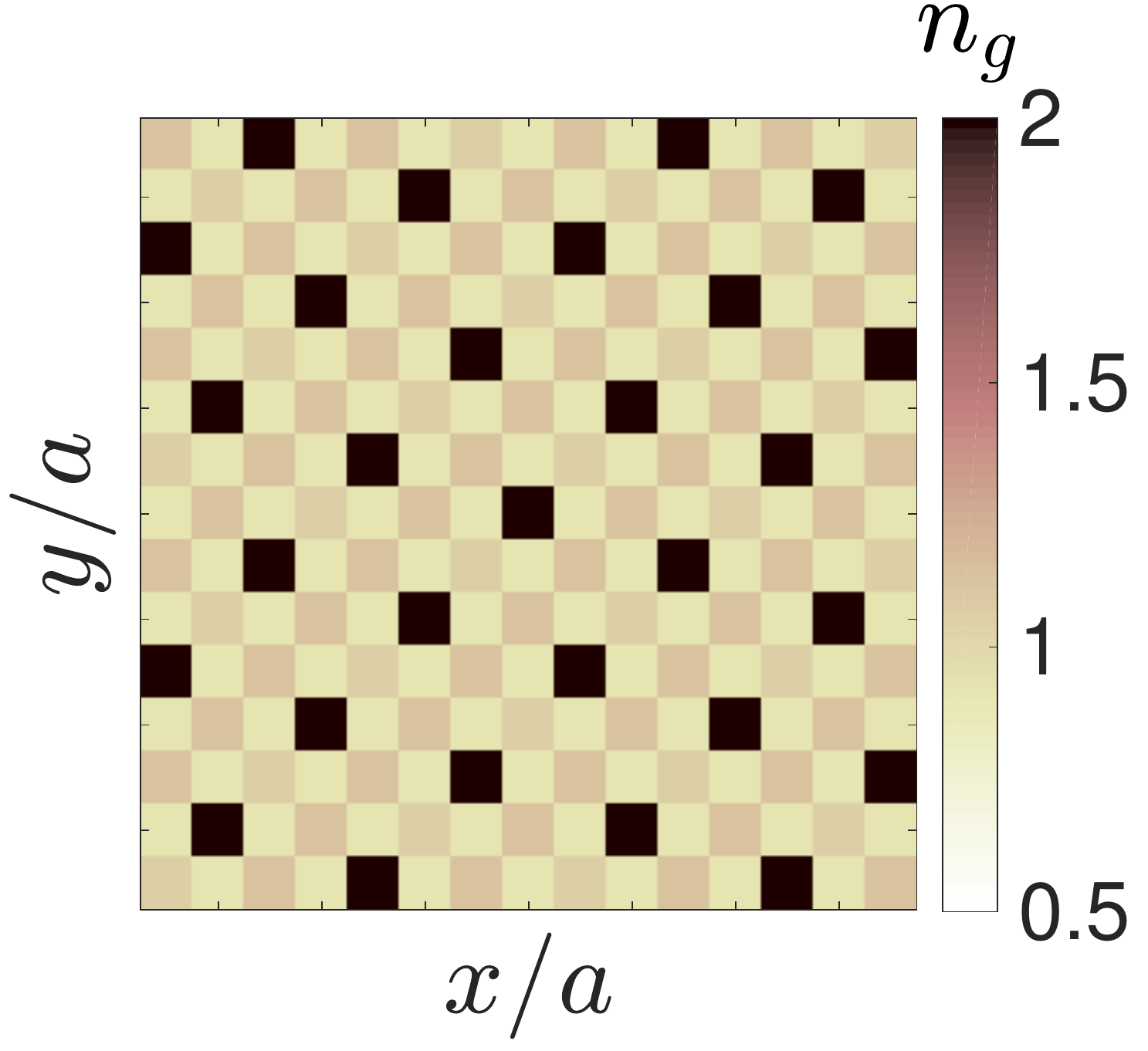}}&
\subfloat[$t = 50 \mu s$]{
\includegraphics[width = .14\textwidth, trim = 10 10 0 0, clip]{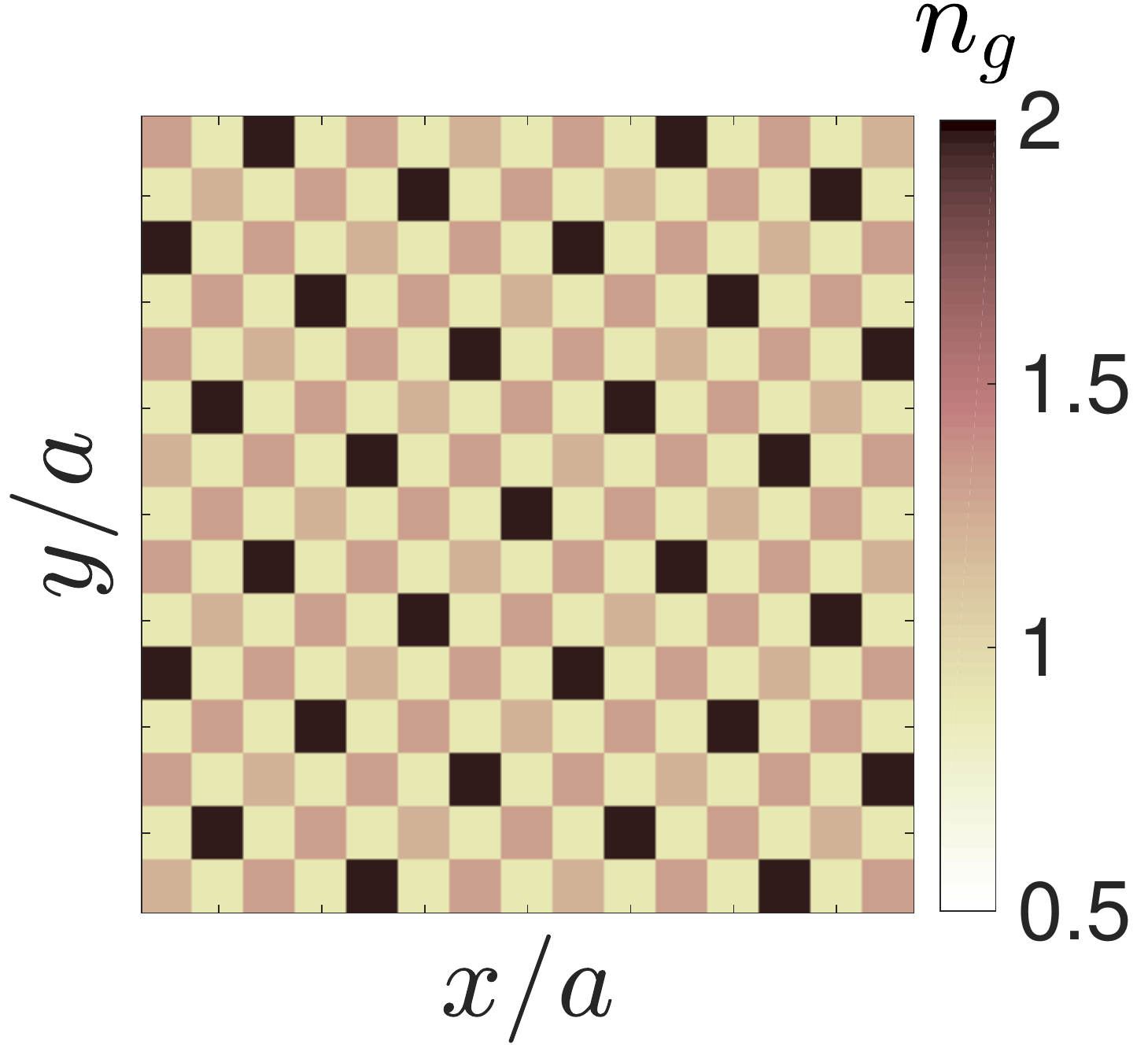}}\\ [-2ex]
\subfloat[$t = 100 \mu s$]{
\includegraphics[width = .14\textwidth, trim = 10 10 0 0, clip]{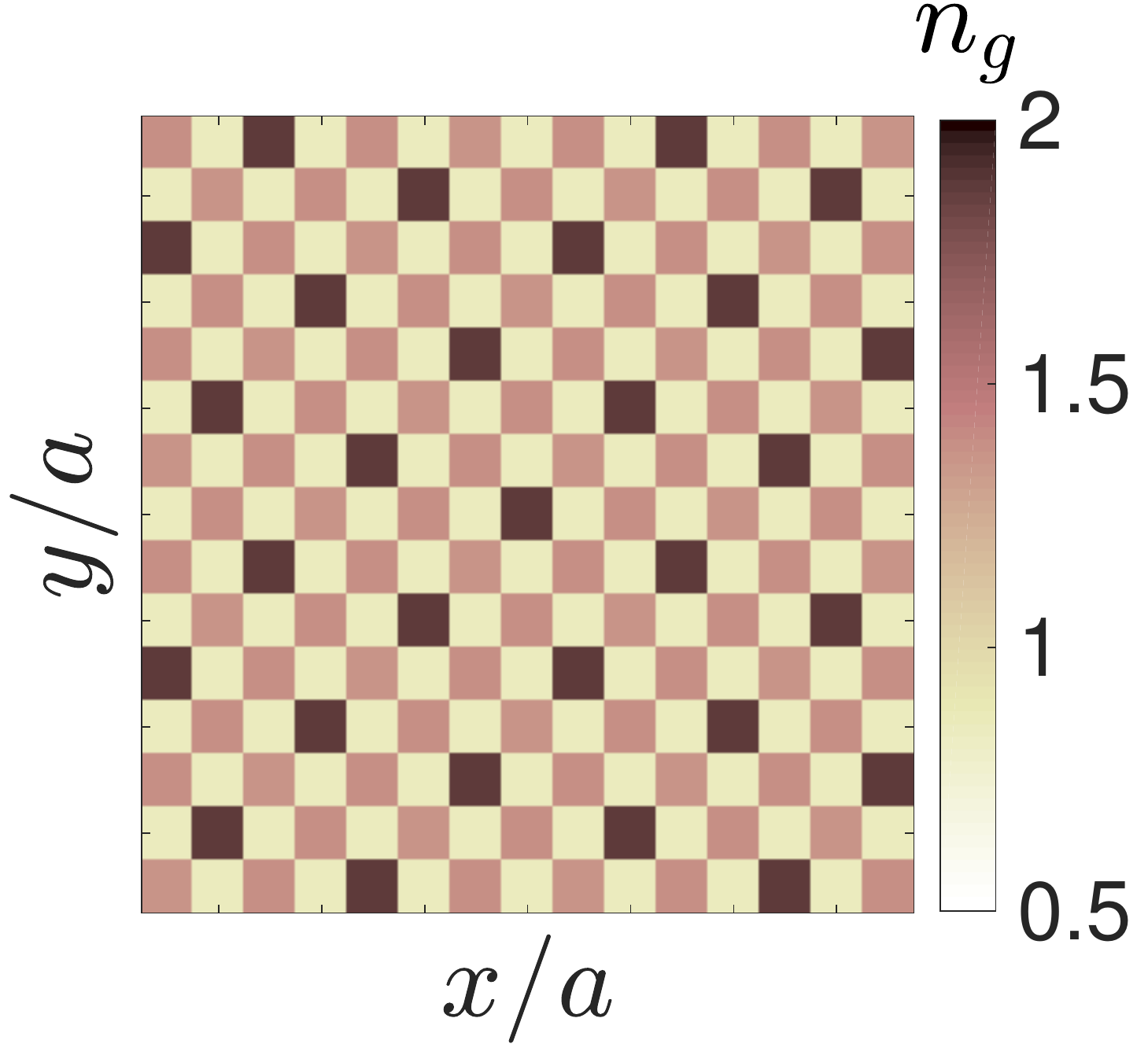}}&
\subfloat[$t = 200 \mu s$]{
\includegraphics[width = .14\textwidth, trim = 10 10 0 0, clip]{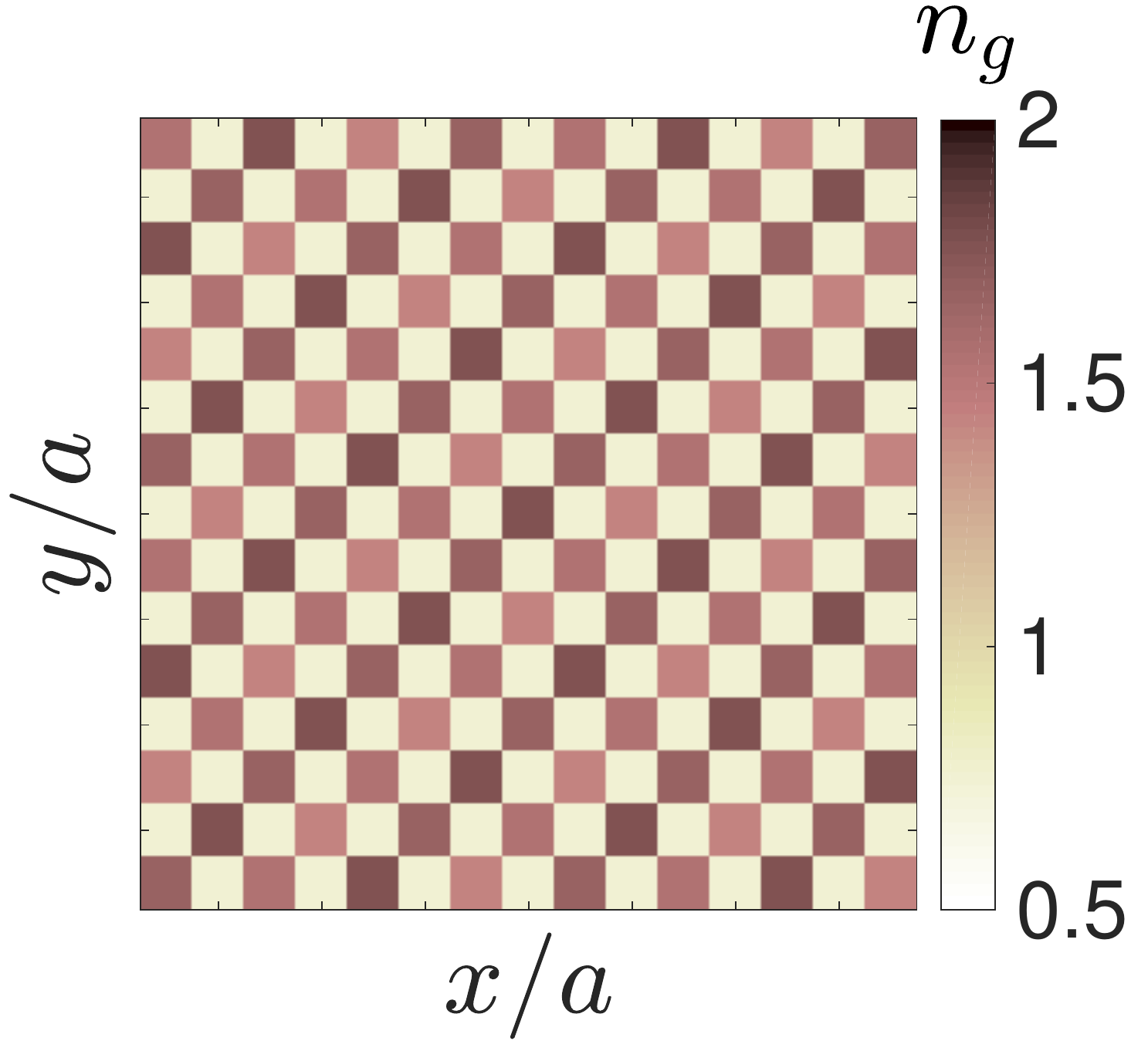}}&
\subfloat[$t = 1000 \mu s$]{
\includegraphics[width = .14\textwidth, trim = 10 10 0 0, clip]{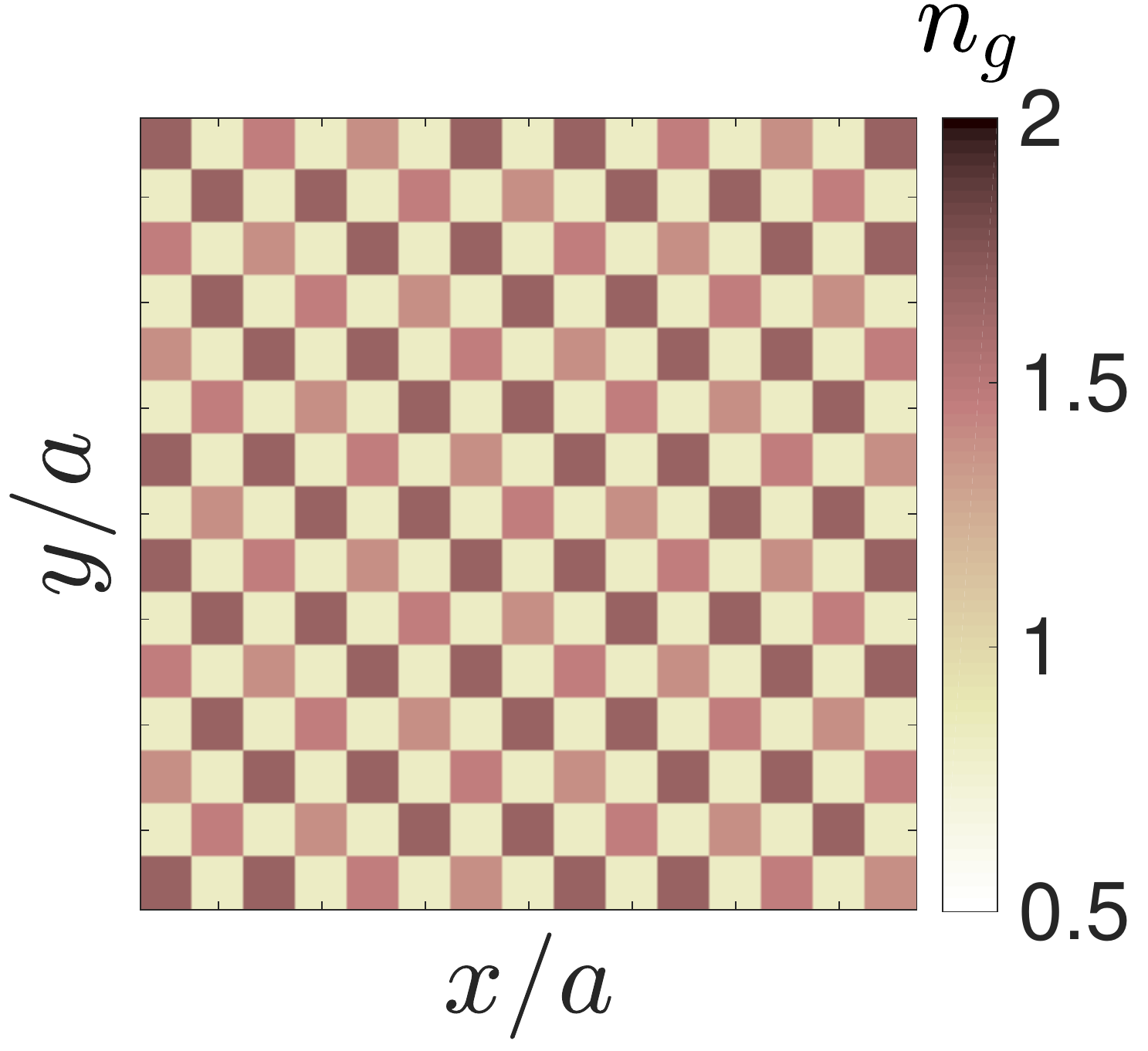}}\\
\end{tabular}
\caption{Time evolution of $n_g$ up to $ 1000 \mu s$ starting with the SS configuration shown in Fig.~\ref{fig:initial-supersolid} for $\Gamma/\Omega = 1$ and $\kappa/\Omega = 0$. Starting with an initial SS, the spontaneous emission process drives the system into a new SS quantum phase with a checkerboard-like long-range order.} \label{fig:transitionSStoCBSS}
\end{figure}
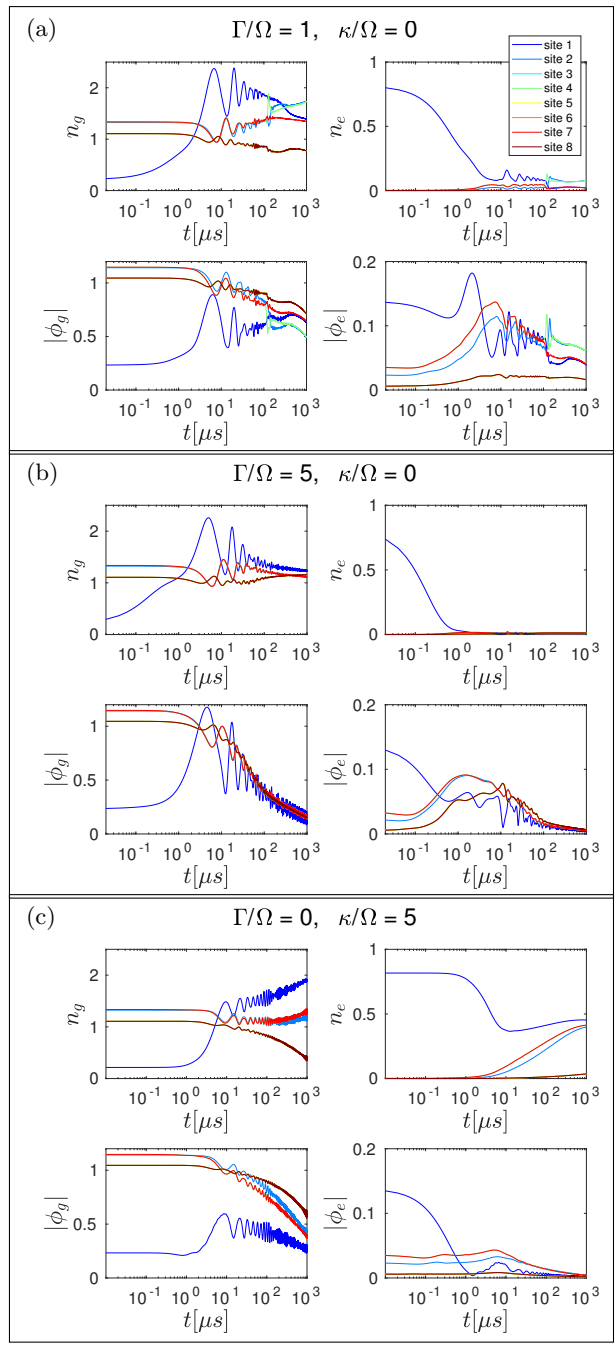
\begin{figure}[t]
\centering
\subfloat{
        \fbox{
        \begin{picture}(219,163)
        \put(0,0){\includegraphics[width=.45\textwidth, trim = {0.5cm 0.2cm 0.5cm 0.5cm}, clip]{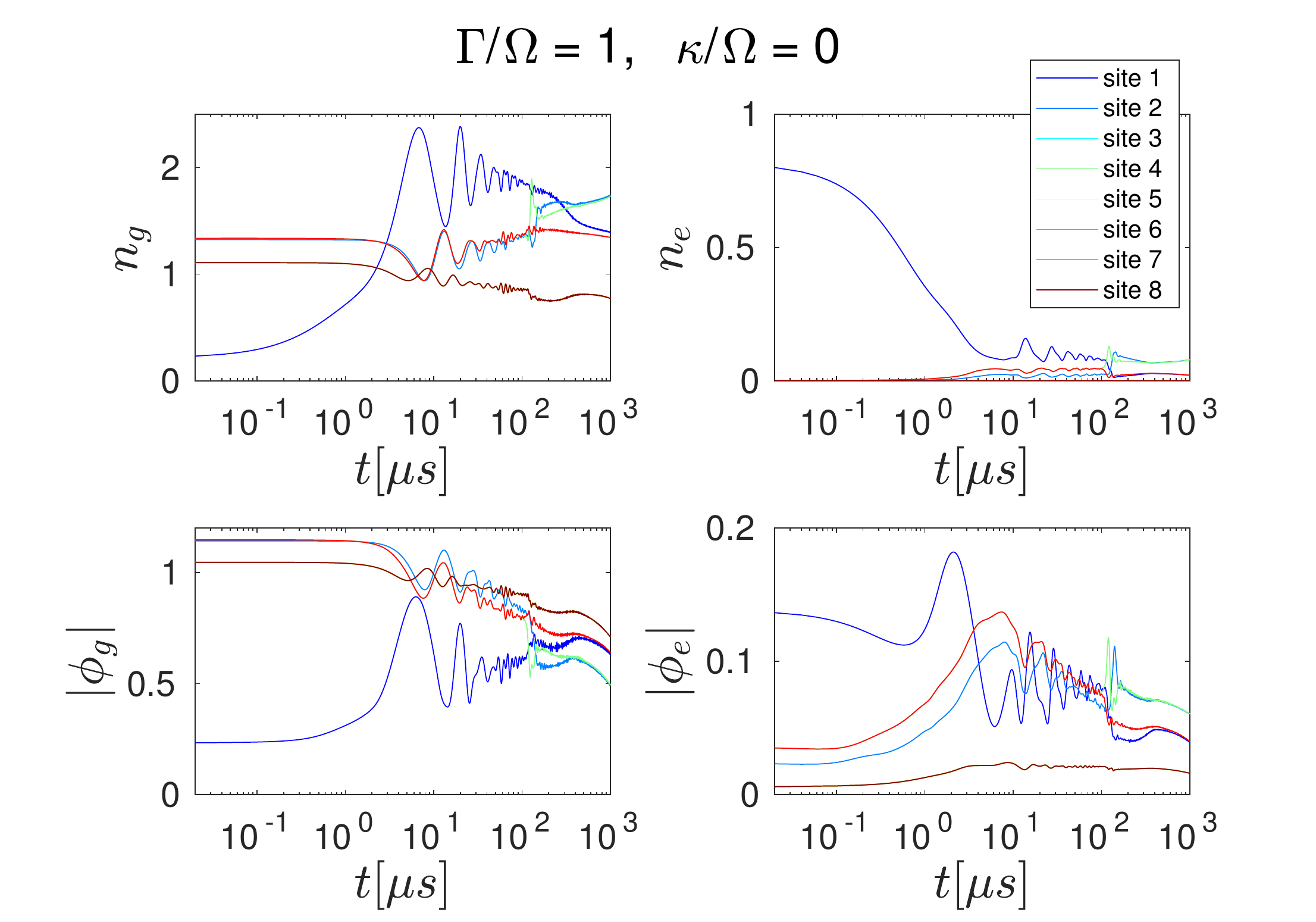}}
        \put(0,155){(a)} 
        \end{picture}}}\\ [-0.7ex]
\subfloat{
        \fbox{
        \begin{picture}(219,163)
        \put(0,0){\includegraphics[width=.45\textwidth, trim = {0.5cm 0.2cm 0.5cm 0.5cm}, clip]{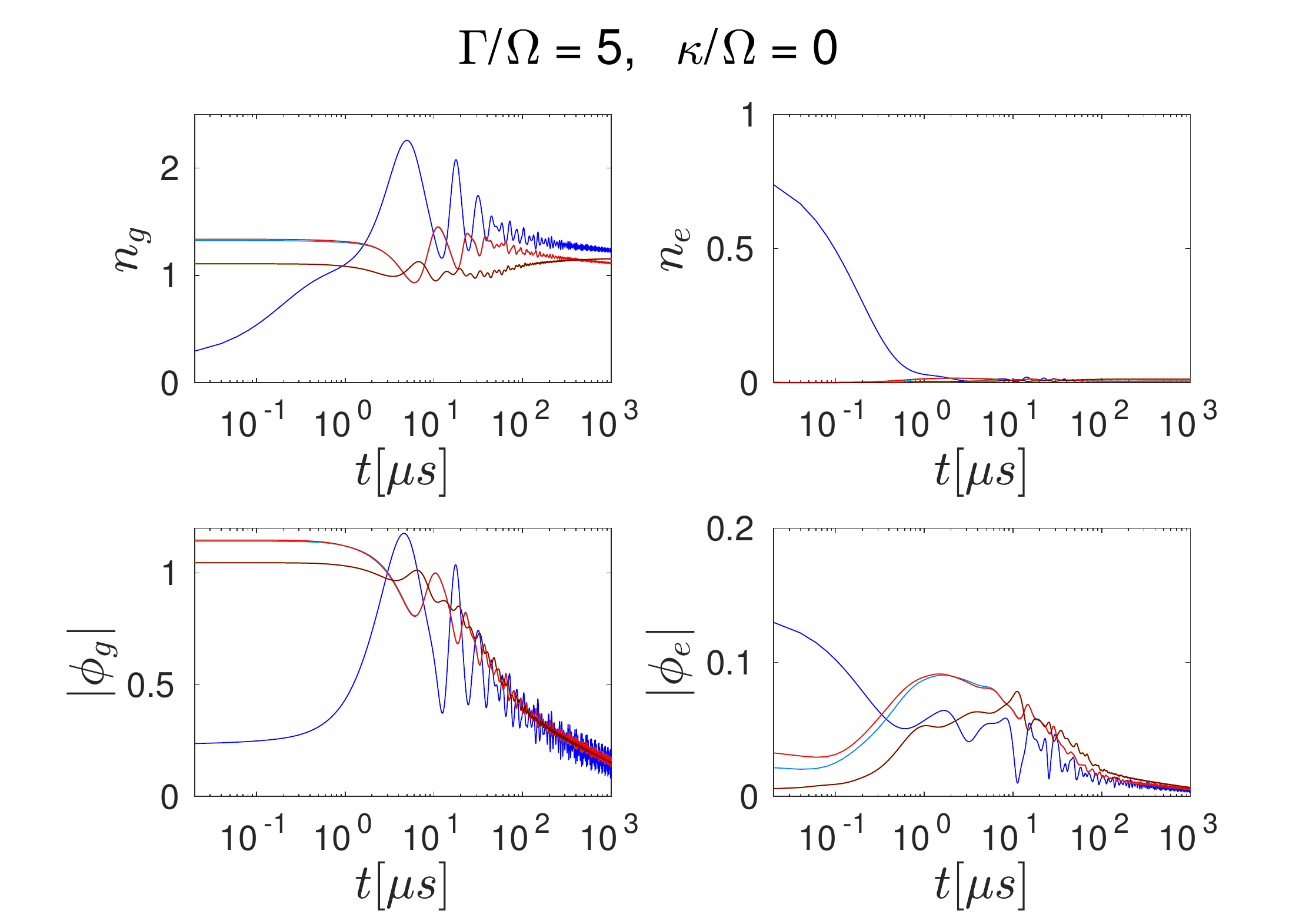}}
        \put(0,155){(b)}
        \end{picture}}}\\ [-0.7ex]
\subfloat{
        \fbox{
        \begin{picture}(219,163)
        \put(0,0){\includegraphics[width=.45\textwidth, trim = {0.5cm 0.2cm 0.5cm 0.5cm}, clip]{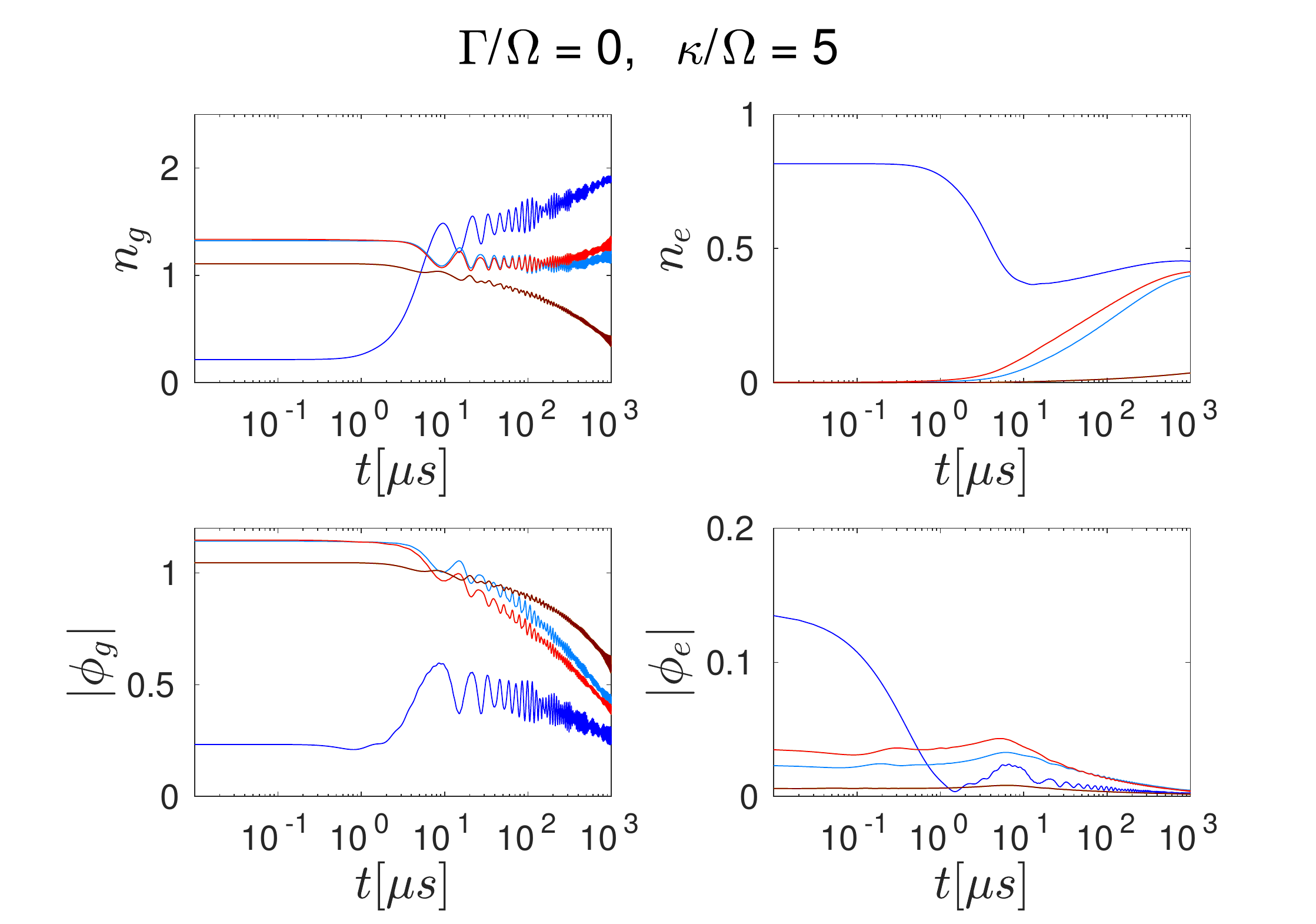}}
        \put(0,155){(c)}
        \end{picture}}}\\
    \caption{Time evolution of relevant observables for an initial SS state shown in Fig. \ref{fig:initial-supersolid} for different decay and dephasing rates $\Gamma/\Omega$ and $\kappa/\Omega$. A low spontaneous emission rate induces a transition into a SS with new crystalline structure, while strong decay lead to the homogenization of the system. Dephasing does not destroy long-range order.} \label{fig:ss-evolution}
\end{figure}

\noindent
We simulate the time evolution of the SS state depicted in Fig. \ref{fig:initial-supersolid} for various sets of decay and dephasing rates. Plotting the dynamics of the observables of each sublattice is not favorable, since their number and shape may vary along time. Instead we plot the dynamics of each site of the primitive cell, which consists of eight sites in total. First, we consider only spontaneous emission ($\Gamma/\Omega \neq 0, \kappa/\Omega = 0$). For small $\Gamma/\Omega$ (see Fig. \ref{fig:ss-evolution}(a)), a finite loss of condensate order of both atomic states is observed. Even though the spontaneous emission rate is small, the Rydberg fraction $n_e$ of the system drops fast. After approximately 150$\mu s$, the attribution of sublattices of the system changes to five and then relaxes into a new crystalline order characterized by three new sublattices.\\
\begin{figure}[t]
\center
\fbox{\includegraphics[width=1\linewidth, trim = {0cm 0.2cm 0.5cm 0.2cm}, clip]{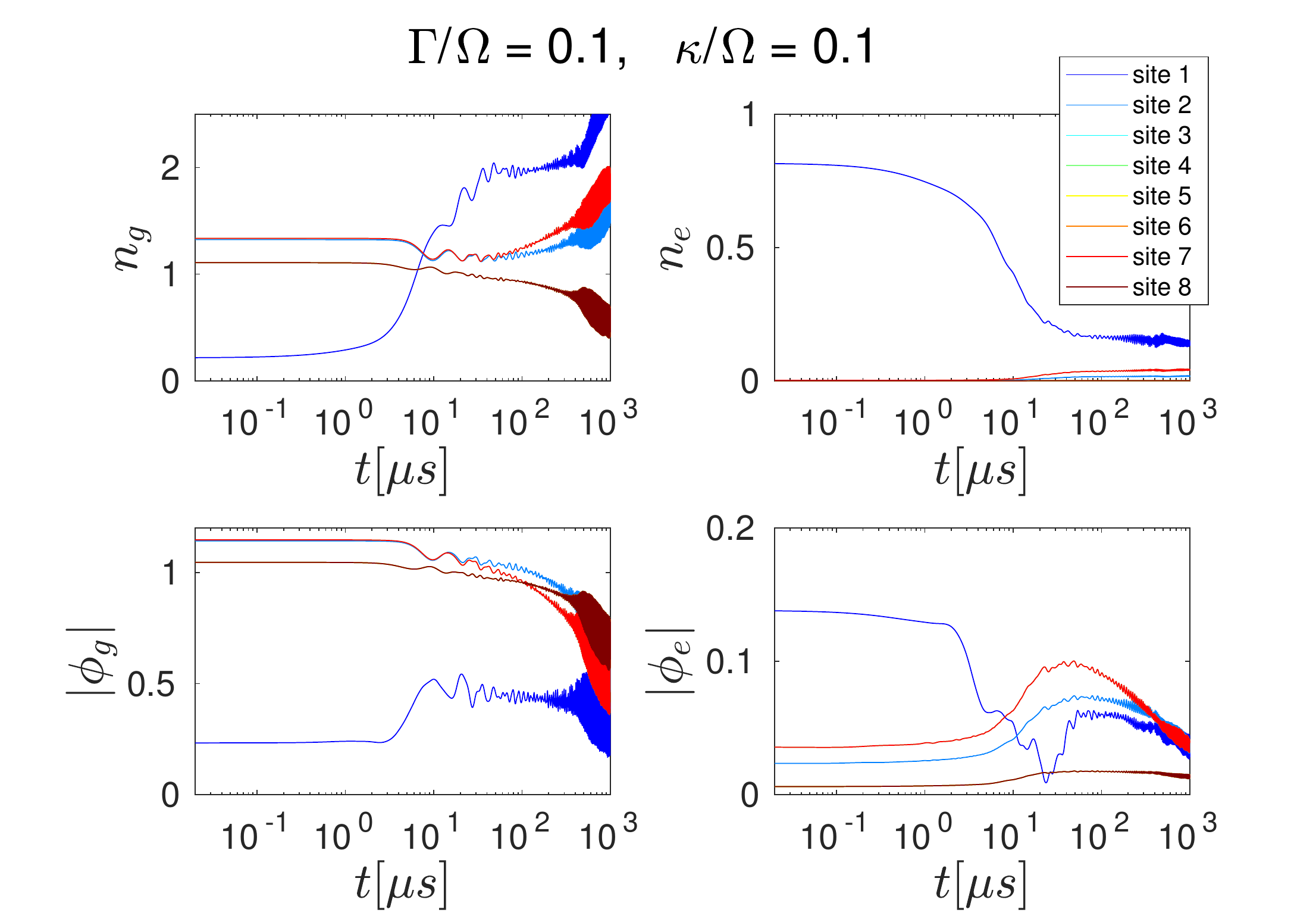}}
\caption{Time evolution of relevant observables for an initial SS state shown in Fig. \ref{fig:initial-supersolid} for fixed rates $\Gamma/\Omega = 0.1$ and $\kappa/\Omega = 0.1$. Fluctuations are strong, but the depletion of the condensate is slow and the system remains a SS for long times.}
\label{fig:realistic}
\end{figure}
In order to gain insight into the time evolution, we depict the real-space distribution of the observable $n_g$ at various times and observe the dynamical phase crossover in Fig. \ref{fig:transitionSStoCBSS}. The system relaxes into a quantum phase with crystalline structure resembling a checkerboard-configuration. The occupation number of the ground and Rydberg state $n_g$ and $n_e$ are seemingly constant in time and both condensate order parameters $\phi_g$ and $\phi_e$ decay at a low rate, rendering the newly obtained checkerboard-like SS quantum phase a long-lived one. On the other hand, high spontaneous emission rates (see Fig. \ref{fig:ss-evolution}(b)) cause a relaxation into a homogeneous state. The rapid decay of all observables persists up to approximately 200$\mu s$. Afterwards, the ground state occupation number $n_g$ remains steady at a finite value, while the condensate order parameter of both states $\phi_g$ and $\phi_e$  further decays slowly. The system loses a significant amount of the Rydberg fraction $n_e$ after a short time due to the high spontaneous emission rate. The vanishing population of the Rydberg state is responsible for the transition into homogeneity, as the system no longer is subject to long-range interaction.Furthermore the attribution of sublattices stays the same in comparison to the case of low spontaneous emission rate.\\
The case of no spontaneous emission and only dephasing ($\Gamma/\Omega = 0, \kappa/\Omega \neq 0$) is depicted in Fig. \ref{fig:ss-evolution}(c). The expectation values initially strongly fluctuate, while all observables show a slow relaxation. Although the system retains the identical four sublattices, their observables evolve over time: The overall condensate order parameter decays and the Rydberg fraction $n_e$ of the system grows after an initial drop.\\
Compared to the case of spontaneous emission only, even high dephasing rates do not seem to lead to homogenization of the system, whereas high spontaneous emission rates do, since dephasing does not induce a loss of Rydberg excitations in the system.\\
\begin{figure}[t]
\center
\includegraphics[width=1.00\linewidth, trim = {1cm 0.8cm 0.1cm 0.2cm}, clip]{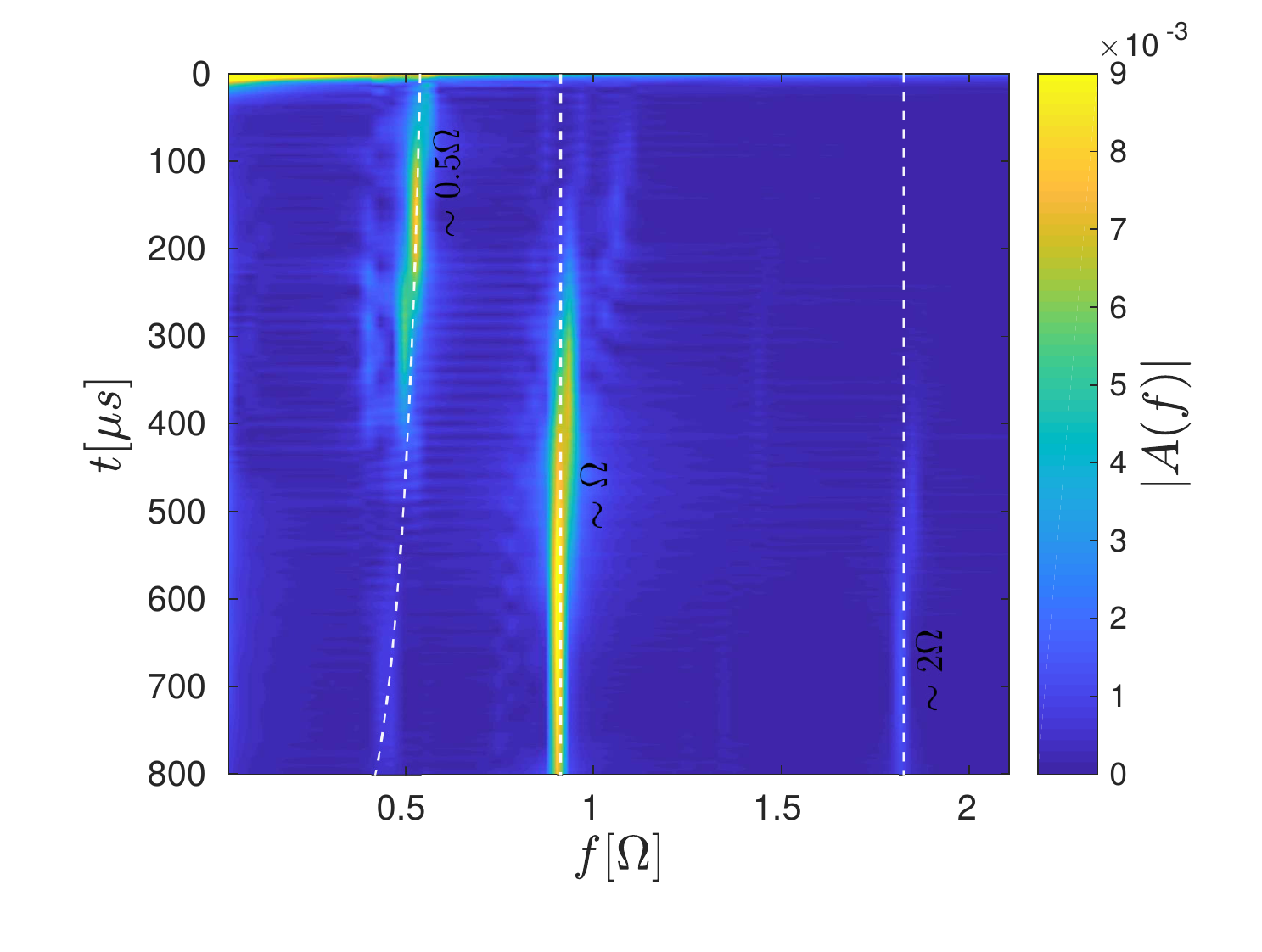}
\caption{Frequency spectrum of the site-averaged time evolution of the ground state occupation number $n_g$ depicted in Fig. \ref{fig:realistic} for various time-windows of length $200 \mu s$ with varying starting time $t$. Dominant frequencies are found around $0.5\Omega$ and $\Omega$ and a small peak is hinted around $2\Omega$.}
\label{fig:fft_mean}
\end{figure}
\par
The time evolution with low but finite decay and dephasing rates ($\Gamma/\Omega \neq 0, \kappa/\Omega \neq 0$) is shown in Fig. \ref{fig:realistic}. On top of the evolution of the observables on each of the four sublattices, strong fluctuations emerge. By performing a Fourier transformation of the time evolution, we clarify the origin of the dynamics. We consider the evolution of the ground state occupation number $n_g$ of Fig. \ref{fig:realistic} and perform a Fourier transformation for several time windows. In Fig. \ref{fig:fft_mean}, we plot the power of the signal versus frequency for different time windows. Two prominent peaks are seen around $0.5 \Omega$ and $\Omega$. While the last one remains fixed throughout the time evolution, the first slightly varies in frequency. Comparing the parameters of the Hamiltonian \eqref{eq:HamiltApprox}, we determine that the first peak belongs to the tunneling process of the ground state atoms. Within the mean-field apprximations used in the derivation of the Hamiltonian, the tunneling rate $t_g$ is scaled with the condensate order parameter of surrounding sites through $\xi^g_i$. Due to the loss of coherence in the system, the condensate order parameter of the ground state $\phi_g$ decays and the effective tunneling rate $t_g \xi^g_i$ along with it. The peak at $\Omega$ represents the Rabi oscillation. A small peak is hinted around $2\Omega$, however, its origin is yet unknown. Furthermore, by looking at the time-dependence of the frequency spectrum, we see that the early dynamics arise from the hopping mechanism, whereas late dynamics are determined by the Rabi oscillations.\\

A Fourier transformation of the dynamics depicted in Fig. \ref{fig:ss-evolution}(a)-(c) yields a highly similar frequency spectrum, although the small $2 \Omega$-peak does not appear.\\
Although those fluctuations are present, the evolution of the SS configuration at later times is slow which renders the observed quantum phase long-lived.\\
\par
We conclude that each dynamical phase crossover is preceded by an initial strong fluctuation of all observables, caused by the sudden enabling of the decay and dephasing processes, inducing a initial perturbation before going into a smooth relaxation. In the presence of low decay or dephasing rates, we observe a crossover into a long-lived transient SS, which is approximately checkerboard-ordered, while high rates of the non-unitary processes lead to a dominant loss of condensate and homogeneity in the case of the spontaneous emission.
\subsection{Detuning-quench time evolution}
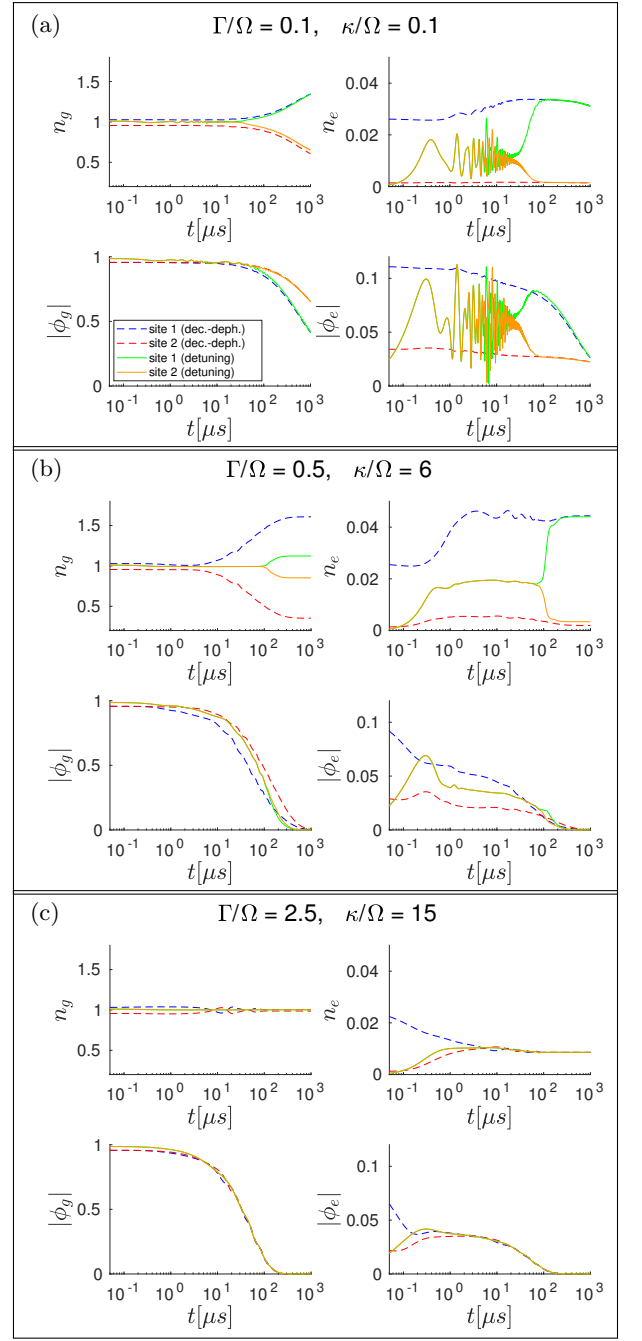
\begin{figure}[t]
\subfloat{
        \fbox{
        \begin{picture}(219,163)
        \put(0,0){\includegraphics[width=.45\textwidth, trim = {0.5cm 0.2cm 0.5cm 0.5cm}, clip]{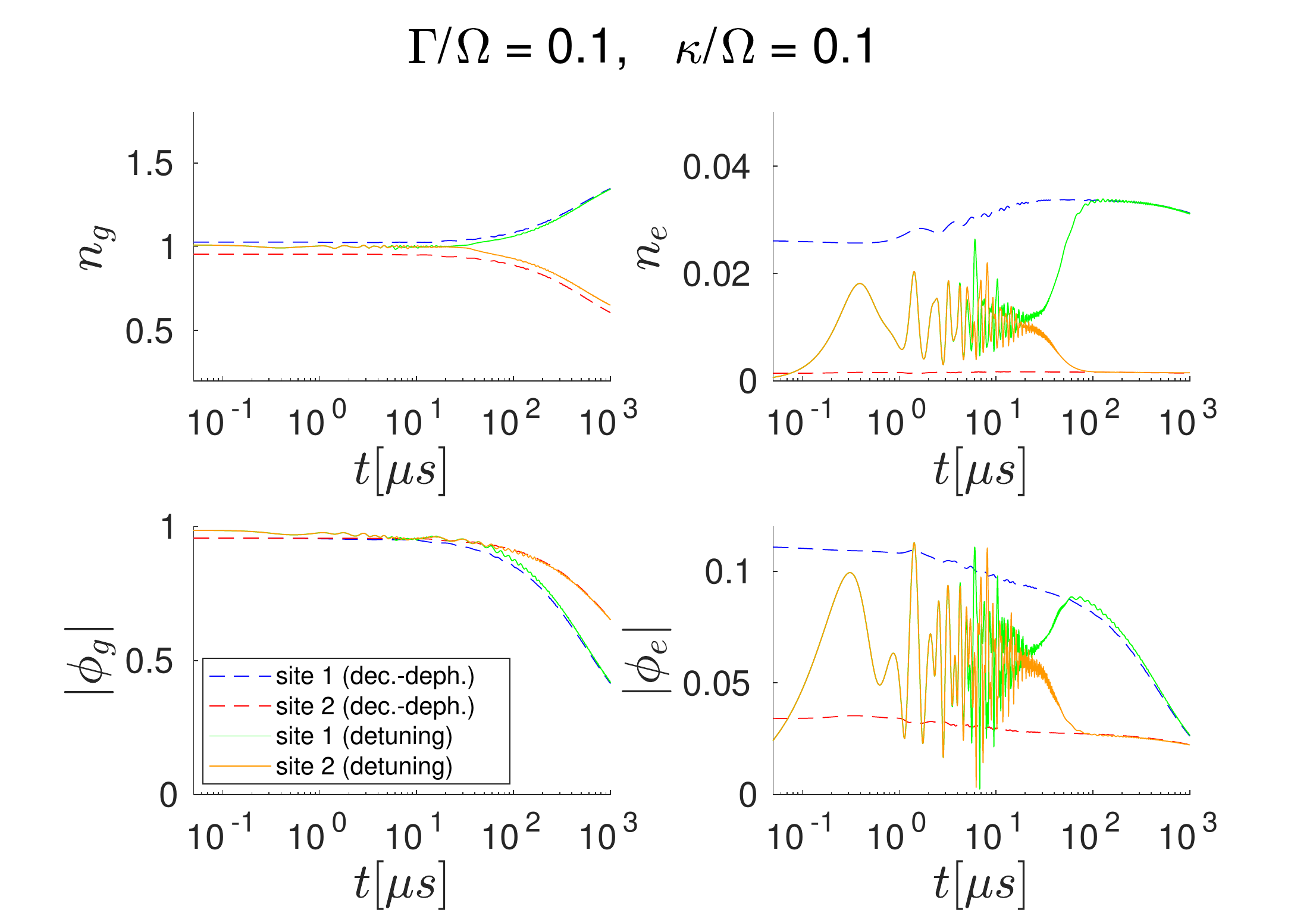}}
        \put(0,155){(a)} 
        \end{picture}}}\\ [-0.7ex]
\subfloat{
        \fbox{
        \begin{picture}(219,163)
        \put(0,0){\includegraphics[width=.45\textwidth, trim = {0.5cm 0.2cm 0.5cm 0.5cm}, clip]{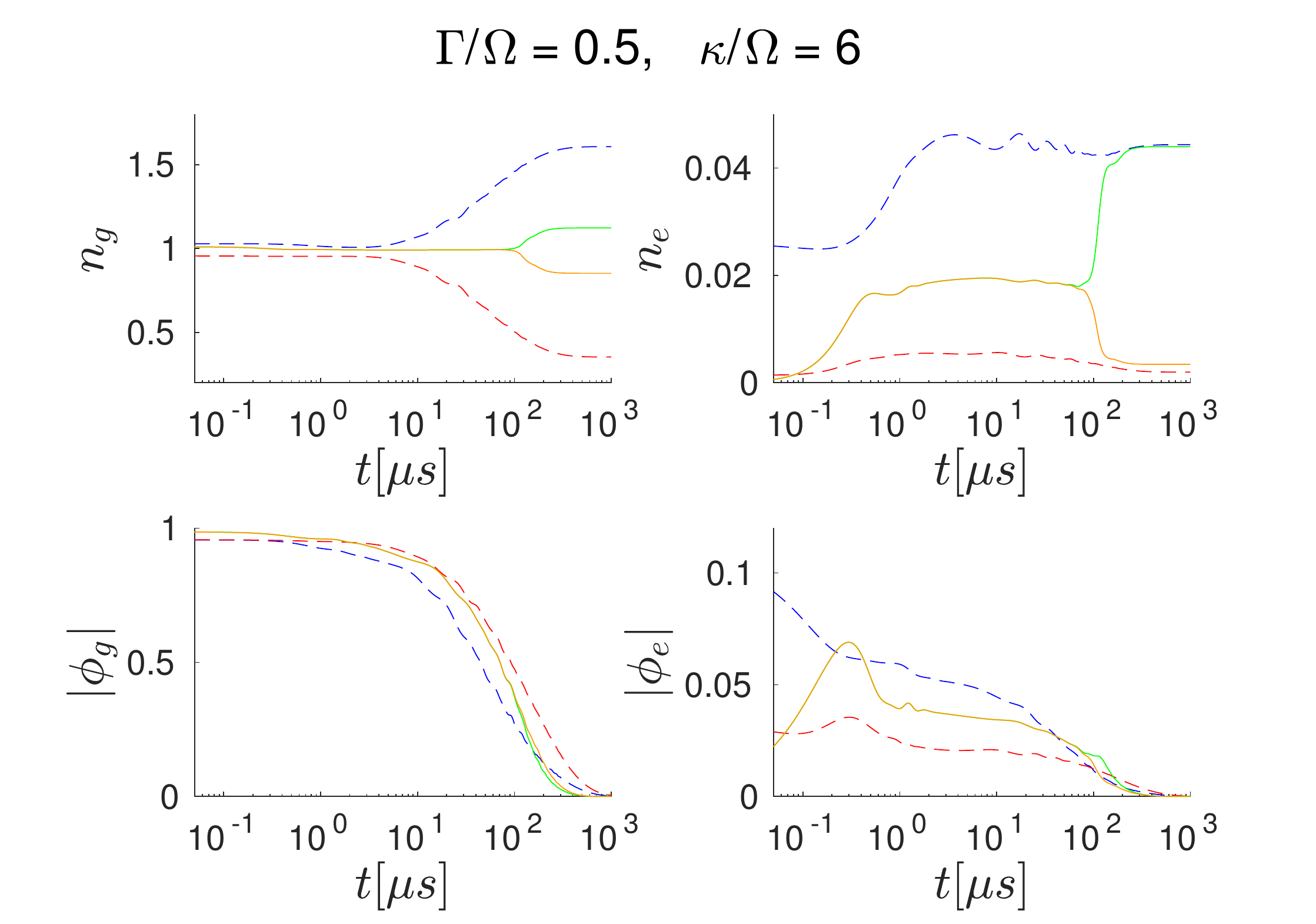}}
        \put(0,155){(b)}
        \end{picture}}}\\ [-0.7ex]
\subfloat{
        \fbox{
        \begin{picture}(219,163)
        \put(0,0){\includegraphics[width=.45\textwidth, trim = {0.5cm 0.2cm 0.5cm 0.5cm}, clip]{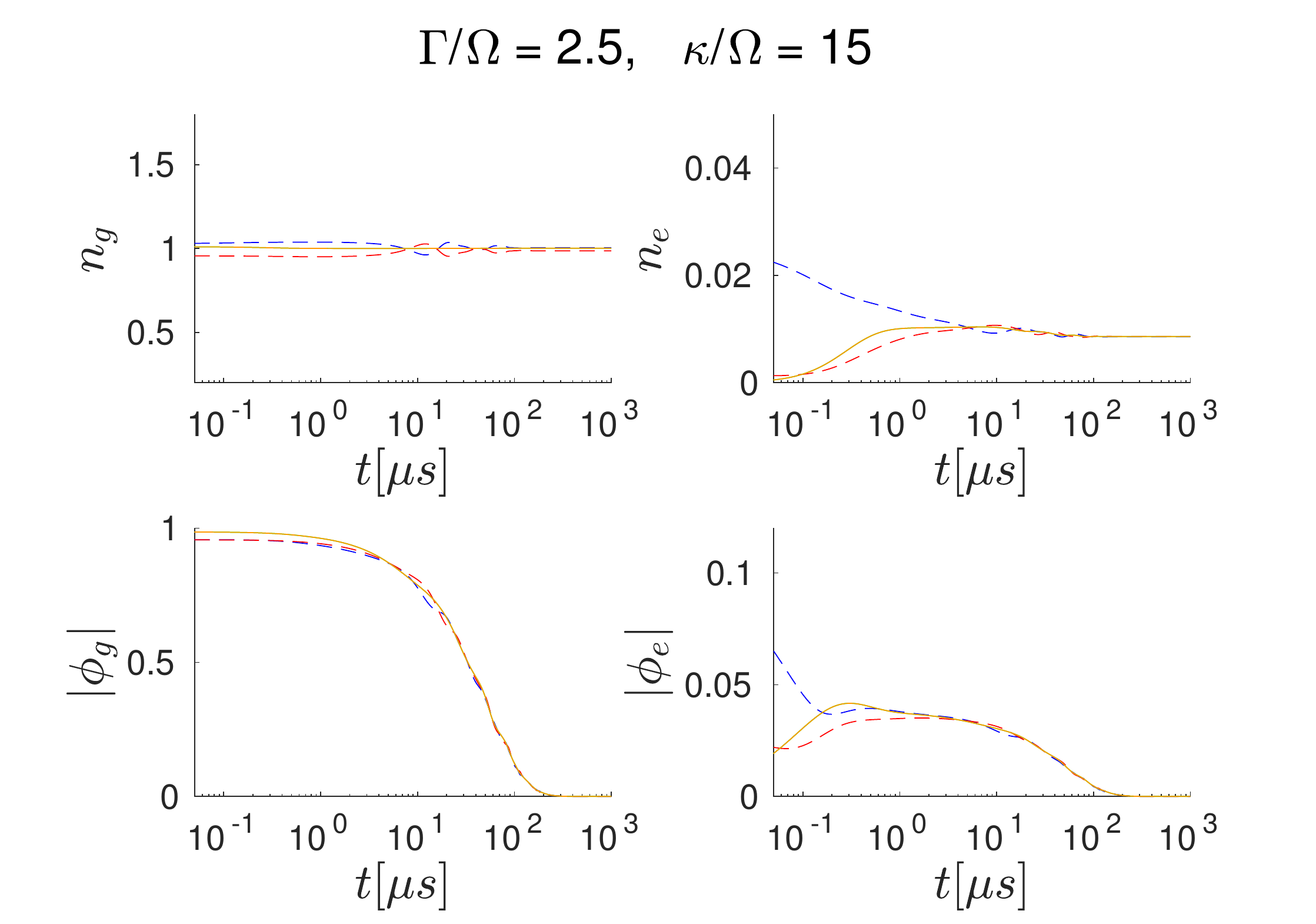}}
        \put(0,155){(c)}
        \end{picture}}}\\ 
    \caption{Comparison between the decay-dephasing-quench time evolution of the CB-SS (dashed blue and red lines from Fig. \ref{fig:cbss-evolution}(a)-(c)) and the detuning-quench time evolution of an initially homogeneous system (solid green and orange lines) for different decay and dephasing rates $\Gamma/\Omega$ and $\kappa/\Omega$.} \label{fig:quench-evolution}
\end{figure}
We now investigate whether the long-lived SS quantum phases discussed previously are dependent on the initial state. If not, we are able to truly refer to those phases as decay-dephasing-induced states, since their dependence on the rates of the non-unitary channels can be considered a generic property. For that purpose, we perform detuning-quench-type time evolution simulations: We start with a homogeneous configuration corresponding to far-negative detuning ($\Delta/\Omega \rightarrow -\infty$), which results in a homogeneous system without Rydberg excitations, and then perform a time evolution after a quantum quench to a suddenly enabled transition to the Rydberg state with finite $\Delta/\Omega$.\\
Using the same parameters as in Fig. \ref{fig:initial-checkerboard} and Fig. \ref{fig:initial-supersolid}, we start with an initial homogeneous state and simulate the time evolution for different decay and dephasing rates. We set the single-site filling of the system to be equal to the average filling of the respective SSs, which we used as initial states in \ref{ch:decay-dephasing}. Since the time evolution preserves the total particle number of the system, we expect that for equal initial average occupation number $\bar{n}$ and equal parameters the homogeneous state might converge to the dynamics previously observed for the decay-dephasing-quenches.\\
\par
\begin{figure}[t]
\center
\fbox{\includegraphics[width=0.95\linewidth, trim = {0.5cm 0.2cm 0cm 0.2cm}, clip]{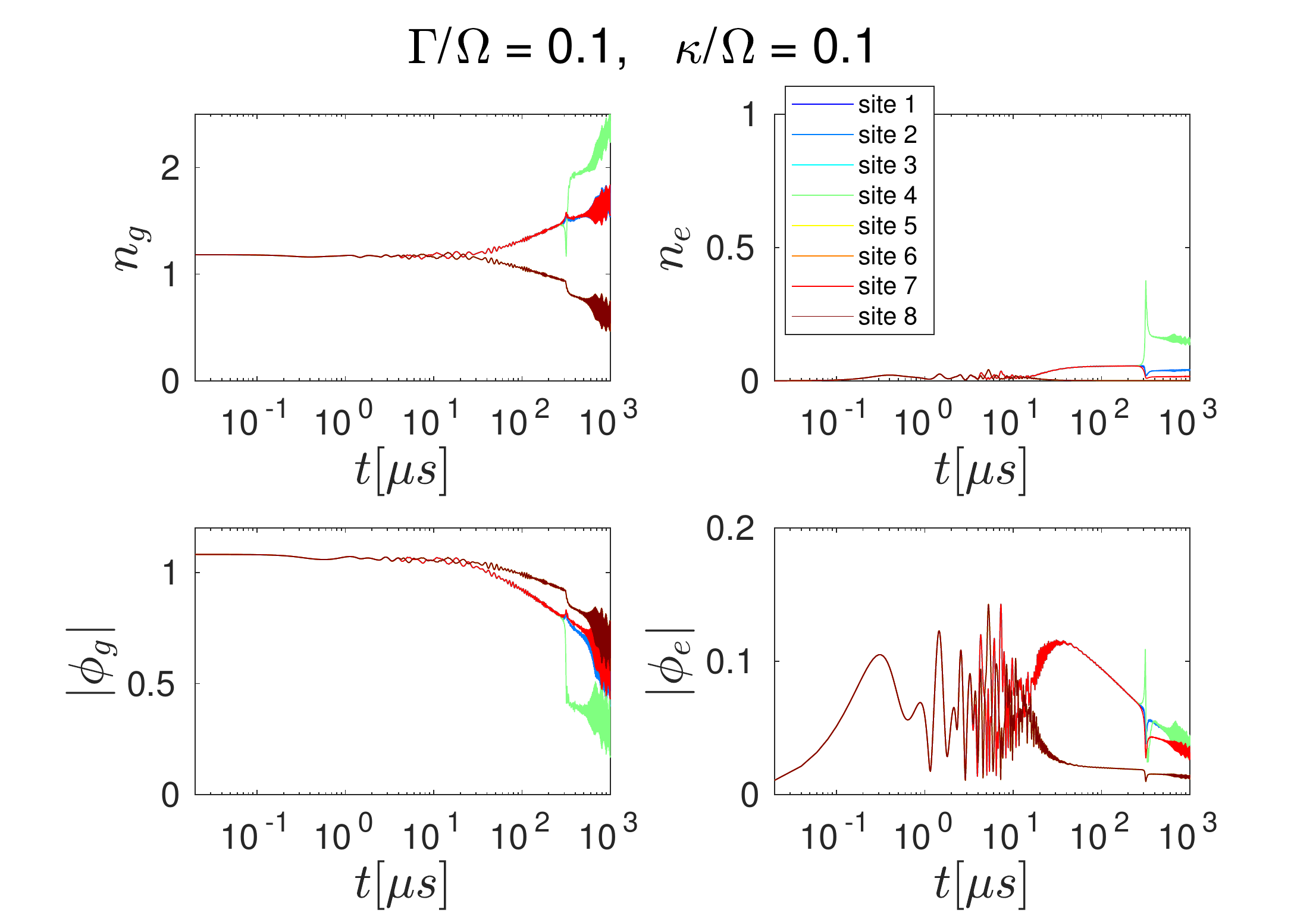}}
\caption{Time evolution of relevant observables for an initial homogeneous state with the parameters used in Fig. \ref{fig:initial-supersolid} for fixed rates $\Gamma/\Omega = 0.1$ and $\kappa/\Omega = 0.1$. The system goes through two symmetry breakings and converges to the evolution of the initial quantum phase with initial supersolid-configuration (compare with Fig. \ref{fig:realistic}).}
\label{fig:quench-ss}
\end{figure}
\par
First, we perform the time evolution of the homogeneous initial state with the average filling and parameters referenced in Fig. \ref{fig:initial-checkerboard} and the decay and dephasing rates used in Fig. \ref{fig:cbss-evolution}. For low rates (see Fig. \ref{fig:quench-evolution}(a)), the detuning-quench of the system induces strong initial fluctuations, which eventually relax to the dynamics of the time evolution of the decay-dephasing quench. We observe that not only the final state, but also the dynamics are primarily determined by the decay and dephasing rates. Intermediate rates suppress the initial fluctuations and delay the formation of a checkerboard (see Fig. \ref{fig:quench-evolution}(b)). Despite the convergence of the ground state occupation number $n_g$ of the initially homogeneous system to different values, the dynamics of all other observables closely resemble the dynamics with initial CB-SS configuration, especially the Rydberg state occupation number $n_e$. In the case of high rates of the non-unitary processes (see Fig. \ref{fig:quench-evolution}(c)), the symmetry-breaking is prevented and the system stays homogeneous. This also indicates that the previously discussed non-vanishing imbalance observed in Fig. \ref{fig:cbss-evolution}(c) is indeed an artifact of the mean-field approximation and should vanish for the parameters, decay and dephasing rates used. The other observables, however, show striking similarities to the decay-dephasing-quench time evolution simulations.\\
We then perform the detuning-quench-type time evolution with the average filling and parameters referenced in Fig. \ref{fig:initial-supersolid}. We depict the dynamics at low decay and dephasing rates $\Gamma/\Omega = 0.1, \kappa/\Omega = 0.1$ for an initially homogeneous system (see Fig. \ref{fig:quench-ss}). After a few microseconds, the system goes through its first symmetry-breaking and develops a checkerboard-configuration with two sublattices. Around $250 \mu s$, the system breaks its symmetry a second time and converges to the time evolution of the decay-dephasing-quench (shown in Fig. \ref{fig:realistic}).\\
\par
We conclude that the long-time dynamics and long-lived states of the system are not dependent on the initial state in the case of low decay and dephasing rates, whereas they may vary for higher rates. Since realistic experimental values of the non-unitary rates are at most $\Gamma_{max}/\Omega = 0.5$ and $\kappa_{max}/\Omega = 0.1$ \cite{MagneticInteraction,DissipationI}, it is safe to assume that the experimental observation of such long-lived SS states is possible.

\section{Conclusions and outlook}
Through the treatment of the many-body Hamiltonian within the Gutzwiller approximation, a variety of equilibrium quantum phases, in particular SSs, have been obtained in experimentally feasible regimes. The successful application of the master equation in Lindblad form allowed us to include the effects of decay and dephasing of the Rydberg state, and lead us to interesting dynamical crossovers in the time evolution. For the studied CB-SS quantum phase, high rates of decay or dephasing result in the homogenization of the system, while low rates preserve the checkerboard structure of the system. Furthermore, the condensate order parameter decreases at a speed that depends on the rates of the spontaneous emission and dephasing. In the case of either slow or high non-unitary rates, the condensate order parameter depletes slowly while the decrease is fastest at intermediate rates. For the considered non-checkerboard SS quantum phase, in the presence of decay and dephasing, the initial SS state is not stable. However, a transition into a long-lived transient checkerboard-like SS phase was observed for experimentally realistic decay and dephasing rates. Similar to the results obtained from the time evolution of an initial CB-SS phase, strong dissipation leads to a loss of long-range order and homogenizes the system. Finally, we have verified that the long-lived SS quantum phase is only weakly dependent on the initial state, and should therefore be realizable in experiments.\\
The branching mechanism to contaminant states briefly mentioned in the theoretical part of \ref{sec:IV} could be a good addition in subsequent investigations. Additionally the generalization of the theoretical decay process description with regard to additional loss processes in two-photon excitation schemes would help to enable more experimental setups to study these systems and therefore presents another future line of work.\\
Nonetheless, our results provide a possible avenue towards the observation of interaction driven supersolid order.\\

\section*{Acknowledgements}
We would like to thank H. Weimer, S. Whitlock and J. Zeiher for insightful discussions. Support by the Deutsche Forschungsgemeinschaft via DFG SPP 1929 GiRyd, DFG HO 2407/8-1, DFG SFB/TR 49 and the high-performance computing center LOEWE-CSC is gratefully acknowledged. 

\bibliographystyle{apsrev4-1}
\bibliography{paper_reflist}

\end{document}